\documentclass[conference]{IEEEtran}
\IEEEoverridecommandlockouts
\usepackage{url}
\usepackage{cite}
\usepackage{amsmath,amssymb,amsfonts}
\usepackage{algorithmic}
\usepackage{graphicx}
\usepackage{textcomp}
\usepackage{xcolor}
\usepackage{multirow}
\usepackage{multicol,lipsum}
\usepackage{booktabs}
\usepackage{mathtools}
\usepackage{cuted}
\usepackage{makecell}
\usepackage{dblfloatfix}
\usepackage[normalem]{ulem}
\useunder{\uline}{\ul}{}

\newcommand{\bftab}{\fontseries{b}\selectfont}
\def\BibTeX{{\rm B\kern-.05em{\sc i\kern-.025em b}\kern-.08em
    T\kern-.1667em\lower.7ex\hbox{E}\kern-.125emX}}

\usepackage{amsmath,amsfonts,bm}

\def\eqref#1{equation~\ref{#1}}
\def\1{\bm{1}}

\def\rmB{{\mathbf{B}}}

\def\vi{{\bm{i}}}

\def\vp{{\bm{p}}}

\def\vs{{\bm{s}}}

\def\vu{{\bm{u}}}

\def\vx{{\bm{x}}}
\def\vy{{\bm{y}}}
\def\vz{{\bm{z}}}

\def\mS{{\bm{S}}}

\DeclareMathAlphabet{\mathsfit}{\encodingdefault}{\sfdefault}{m}{sl}
\SetMathAlphabet{\mathsfit}{bold}{\encodingdefault}{\sfdefault}{bx}{n}

\def\gD{{\mathcal{D}}}

\def\gI{{\mathcal{I}}}

\def\sI{{\mathbb{I}}}

\def\sR{{\mathbb{R}}}

\def\sU{{\mathbb{U}}}

\def\sX{{\mathbb{X}}}
\def\sY{{\mathbb{Y}}}
\def\sZ{{\mathbb{Z}}}

\newcommand{\ptrain}{\hat{p}_{\rm{data}}}

\newcommand{\pmodel}{p_{\rm{model}}}

\newcommand{\sigmoid}{\sigma}

\newcommand{\ie}{\textit{i}.\textit{e}.}
\newcommand{\eg}{\textit{e}.\textit{g}.}

\begin{document}
\title{%TotalRecall: A Unified User-Item Matching Framework for Multiple Marketing Purposes of Merchants \\
%UniMatch: A Unified User-Item Matching Framework for the Marketing of Merchants 
%\\on Cloud
UniMatch: A Unified User-Item Matching Framework for the Multi-purpose 
\\Merchant Marketing}
%\author{Anonymous}
%\institute{Anonymous}
% qianchuan.lty@alibaba-inc.com
% dmmeng.dm@alibaba-inc.com
% jiangyu.jiangyu@alibaba-inc.com
% yuyang.sqh@alibaba-inc.com
% zhongyao.wangzy@alibaba-inc.com
% liuhong.liu@alibaba-inc.com
% huan.xu@alibaba-inc.com

\author{    
\IEEEauthorblockN{
    	Qifang Zhao, 
        Tianyu Li,
        Meng Du,
        Yu Jiang,
        Qinghui Sun,
        Zhongyao Wang,
        Hong Liu,
        Huan Xu
     }
    \IEEEauthorblockA{
       % \textit{Data Technology} \\
       \textit{Alibaba Group}, Hangzhou, China \\
       \{james.zqf, qianchuan.lty, dmmeng.dm, jiangyu.jiangyu\}@alibaba-inc.com\\
       \{yuyang.sqh, zhongyao.wangzy, liuhong.liu, huan.xu\}@alibaba-inc.com\\
    }
}

\iffalse
\author{Anonymous}
\institute{Anonymous}
\author{First Author\inst{1}\orcidID{0000-1111-2222-3333} \and
Second Author\inst{2,3}\orcidID{1111-2222-3333-4444} \and
Third Author\inst{3}\orcidID{2222--3333-4444-5555}}
%
\authorrunning{F. Author et al.}
\institute{Princeton University, Princeton NJ 08544, USA \and
Springer Heidelberg, Tiergartenstr. 17, 69121 Heidelberg, Germany
\email{lncs@springer.com}\\
\url{http://www.springer.com/gp/computer-science/lncs} \and
ABC Institute, Rupert-Karls-University Heidelberg, Heidelberg, Germany\\
\email{\{abc,lncs\}@uni-heidelberg.de}
}
\fi

%
\maketitle              % typeset the header of the contribution
% To utilize the cloud services for intelligent marketing, 
\begin{abstract}

When doing private domain marketing with cloud services, the merchants usually have to purchase different machine learning models for the multiple marketing purposes, leading to a very high cost.
We present a unified user-item matching framework to simultaneously conduct item recommendation and user targeting with just one model. 
We empirically demonstrate that the above concurrent modeling is viable via modeling the user-item interaction matrix with the multinomial distribution, and propose a bidirectional bias-corrected NCE loss for the implementation.
The proposed loss function guides the model to learn the user-item joint probability $p(u,i)$ instead of the conditional probability $p(i|u)$ or $p(u|i)$ through correcting both the users and items' biases caused by the in-batch negative sampling.
In addition, our framework is model-agnostic enabling a flexible adaptation of different model architectures.
Extensive experiments demonstrate that our framework results in significant performance gains in comparison with the state-of-the-art methods, with greatly reduced cost on computing resources and daily maintenance.

% Further, it is able to correct the biases caused by the in-batch negative sampling.

%predicting next-$n$-day behavior %instead of conventional next-item prediction, in order not 
%can achieve improved performance over state-of-the-art (SOTA) results in comparison with previous methods, 
%It is capable of serving for multiple marketing purposes, such as item recommendation and user targeting, simultaneously with just one model.
% with purpose-specific models
%In the internet era, merchants are moving their sales data to the cloud to save cost, for example, Amazon Web Services or Alibaba Cloud. They usually need to buy many machine-learning models for different marketing purposes, which pushes up the cost a lot. To solve the problem, we propose a bidirectional user-item matching framework. It can serve for those purposes with just one model through modeling the user-item joint probability $p(u,i)$ instead of the conditional probability $p(i|u)$ or $p(u|i)$. Extensive experiments show that our model can archive SOTA results. Compare to the alternative methods, our framework can greatly reduce the cost of calculation and maintenance. By letting merchants be affordable for the marketing powered by artifical intellegence (AI), we democratize AI, and contribute to the carbon neutralization at the same time.

\end{abstract}
\section{Introduction}
\label{sec:intro}

Nowadays, merchants commonly sell their products in multiple channels, such as the public platforms like Amazon, Alibaba, and the private channels like their own websites, offline shops, and exclusive customer groups on social medias like Wechat, etc. 
%Marketing in public platforms has been well managed by those e-commerce companies, so merchants are paying more attention to operate their business in the private channels, \ie, private domain marketing. 
The marketing on those public platforms, managed by the e-commerce companies, has reached a limit in recent years.
As a result, merchants are paying more attention to operate their businesses via the private channels, \ie, conducting the private domain marketing.
%They are moving their systems and data from local data centers to the cloud, for example, Amazon Web Services or Alibaba Cloud
In order to manage businesses more effectively, merchants utilize the cloud services like Amazon Web Services and Alibaba Cloud, to link all the private channels. 
%As a result, integrated marketing strategies can be adopted.
% In order to manage the data from different private channels and save the cost, they  are moving their systems and data from local data centers to the cloud, for example, Amazon Web Services or Alibaba Cloud.

%In the cloud, besides the basic data management functionality, merchants also want to utilize the modern machine learning techniques to do intelligent marketing.
The cloud services not only manage data for merchants, but also provide machine learning techniques for the intelligent marketing.
%There are two common marketing strategies of opposite directions: item recommendation~\cite{ricci2011introduction} ($u2i$) and user targeting ($i2u$). 
There are two common marketing directions of the merchants: the item recommendation (IR) \cite{ricci2011introduction} and the user targeting (UT). 
To be more specific, merchants try to keep their high-value users active and loyal by periodically sending them messages or emails with recommended items.
%In user targeting ({\bf UT}), merchants mine the potential buyers of the picked items, e.g., new items or unsalable items, and then send the specific promotion messages to the targeted users. 
Meanwhile, merchants always look forward to discovering the potential buyers for certain items, \eg, new releases or popular products, etc. Then, they can send personalized promotion content to those targeted users.
Owing to the machine learning techniques, both item recommendation and user targeting contribute to the profit of merchants significantly.
%With the help of machine learning, item recommendation and {\bf UT} can earn much more profit. %The other strategy is to recommend personalized items to users.  For convenience, we name the strategies as  and recommendation.
% The AI-aided advertisement will save lots of cost. 

However, merchants have to purchase a handful of machine learning models for different marketing purposes. 
First, the item recommendation usually requires one model. 
Then, the user targeting usually requires more than one model, because practitioners need to create multiple targeting lists according to different promotion subjects, \eg, popular products or bundles of items.
%because practitioners need to build one model for each item or each group of similar items.
It takes great efforts to conduct feature engineering, model training and inference for each model.
%Every model has its own feature engineering, training and inference. 
%Also, the optimization of feature sets and models are error-prone, especially with certain features being shared across different models. 
These practices push up the cost dramatically.

%For example, they plan to find out users who might be interested in newly launched items, and then start a campaign to reach them with sms or email. However, they have to buy different models for various purposes. Assume that we have three tasks abstracted from those purposes, to mine the potential users in the next month, and potential users of picked item, and to recommend items for selected users. We will have to train 3 models with different features, and then predict the results separately. 

%This paper proposes a unified user-item matching framework, named \emph{UniMatch}, to serve the various marketing purposes with one model only.
This paper proposes a unified user-item matching framework, named \emph{UniMatch}, to serve for the item recommendation and user targeting with one model only.
%Previously in recommendation, the conditional probability $p(i|u)$ is the modeling objective~\cite{covington2016deep,li2019multi}. In {\bf UT}, the objective is $p(u|i)$. 
The previous recommendation algorithms utilize the conditional probability $p(i|u)$ as the modeling objective~\cite{covington2016deep,li2019multi}, while the user targeting models are commonly optimized via the objective $p(u|i)$.
%The apparent difference suggests their results cannot be applied interchangeably. In this paper, we propose a framework called UniMatch to serve the two purposes together with one model only. 
In our UniMatch framework, the modeling objective is the joint probability $p(u,i)$, which is implemented with a bidirectional bias-corrected NCE loss, named \emph{bbcNCE}. 
When applied for the item recommendation, $p(u,i)=p(i|u)p(u)$ will produce a similar item list compared to $p(i|u)$ given a specific user. 
The same logic holds for the user targeting as well.
%We propose the bidirectional InfoNCE~\cite{oord2018representation} (bbcNCE) loss and show that it models the jointly probability $p(u,i)$. 
%We theoretically and experimentally show that it can be applied on both recommendation and {\bf UT} after training. 
%Thus, our framework reduces the cost of computing resources and data storage by at least half, and relieves even more burden of the daily maintenance. 
Thus, our framework is able to reduce the cost of computing resources and data storage, and relieve the burden of the daily maintenance as well. 
%One typical objective is to recommend items to users, and the recommender system (IR)~\cite{ricci2011introduction} is developed to solve for it. The other objective is to discover the potential users of the given items, and the send the advertising campaign messages. The potential users mining system ({\bf PUMS}) can be utilized to solve it.

%Different from the real-time recommendation in the e-commerce platforms, merchants usually do the recommendation and {\bf UT} every several weeks or 1 month.
Different from the online recommendation on the e-commerce platforms, merchants usually apply these intelligent marketing models less frequently when doing private domain marketing.
For instance, they send promotion emails or personalized messages weekly or even longer.
%To adapt for this specific scenario, we predict all the items purchased in the next $n$ days (next-$n$-day prediction), instead of predicting the next item (next-item prediction). 
To adapt for this specific scenario, both the potential user and recommended item lists are produced under a next-$n$-day prediction setting. 
%We discover that next-$n$-day prediction gives much better results. Also, we adopt the incremental training schema to feed the data into model according to occurring time.
%We show that it greatly improves the results, because it allows the model to gradually adapt to the distributions close to the target distribution.
%We show that it greatly improves the model performance, because it allows the model to gradually adapt to the distributions close to the target distribution.
Conventionally, both the item recommendation and user targeting tasks are solved by modeling the Bernoulli or multinomial distribution on the user-item interaction matrix.
In this paper, we first theoretically prove that it is equivalent to model with the Bernoulli and multinomial distribution since they converge to the same optima in practice.
Then, we uncover that modeling with the multinomial distribution has better efficiency in terms of the data preparation and model convergence.
Therefore, we follow our discovery and propose a bidirectional NCE loss with bias correction to model the user-item joint probability $p(u,i)$.
Additionally, our framework adopts a classical two-tower architecture which enables a flexible utilization of different models.

Our framework has been implemented in the Alibaba cloud product, \emph{QuickAudience(QA)}\footnote{\url{https://help.aliyun.com/document_detail/136924.html}}, for the intelligent marketing of the merchants. Our contributions are summarized as follows:

\begin{itemize}

\item We present a unified user-item matching framework, \emph{UniMatch}, which trains only one model to serve both the item recommendation and user targeting simultaneously. To the best of our knowledge, this is the first work on the topic.

\item We theoretically prove the equivalence between modeling the user-item interaction matrix with the Bernoulli and multinomial distributions, and empirically demonstrate that modeling with the multinomial distribution yields more robust results with much less resources.
%Its modeling objective is the joint probability of $u$ and $i$.
%It implements the bbcNCE loss and trains models with the joint probability of $u$ and $i$ being the learning objective.
%Practically it is very beneficial to merchants because it can reduce the storage by half and ease the efforts of maintenance a lot.
%(bbcNCE??)

% \item \textbf{next-$n$-day prediction}: We choose the next-$n$-day prediction instead of next-item prediction when preparing the training data. This helps boost the HitRate and Recall a lot. [to be specific!]

\item We propose a bidirectional bias-corrected NCE loss, \emph{bbcNCE}, and train models with the joint probability of $u$ and $i$ being the learning objective in theory. Also, we empirically show that the bbcNCE loss will guide the model to learn the joint distribution. % can produce SOTA results in both the IR and UT, which serves as a proof of our theoretical claim. %

% \item We also propose another realization of learning the joint-probability $p(u,i)$, which models the user-item interaction matrix with Bernoulli distribution. It is implemented using the binary cross entropy loss and specific negative sampling strategy. We theoretically prove it and verify with experiments.

% \item We apply the incremental training procedure and apply the next-$n$-day prediction to fit the private domain marketing scenario on cloud.
% It allows the model training not to learn from the highly biased user-item distributions.

\item Extensive experiments on two public datasets and two real-world datasets demonstrate that the proposed framework consistently yields improved performance, in comparison with the state-of-the-art methods on both item recommendation and user targeting tasks. In addition, our framework saves up to 94\%+ of the total cost compared to previous practices.

%\item \textbf{Achieving results comparable to SOTA in both IR and PUMS}: We show that our models can achieve SOTA results in the applicable scenarios. Specifically, TR is comparable to MF in IR and PUMS. [to be modified!]

\end{itemize}

\section{Preliminaries}
\label{sec:preliminaries}
In this section, we first describe the characteristics of the private domain marketing for merchants, and then introduce how the previous research solves the tasks of IR and UT via modeling the user-item interaction matrix separately with the Bernoulli or multinomial distributions.
%and introduce how we formulate it to a machine learning problem.

\subsection{Problem Definition}
\label{sec:prob-def}

For the private domain marketing, %different from the online recommendation on the public platforms, 
the merchants generally carry out the IR and UT for marketing periodically via their private channels.
%In general, the merchants carry out the private domain marketing activities periodically.
They send out messages or emails to their active users or potential customers, and then expect them to take actions, \eg, visiting offline shops, making inquiries online, and placing an order, etc, some time later.
It relatively takes longer time for the merchants to achieve the private domain marketing results.
%For example, one day they send out messages or emails to users, and then expect them to make a purchase some days later, and it could 1 week, 2 weeks or even months.
%Bearing in mind of the actual application, we formulate the problem in a way different from conventional next-item prediction.
%To fit for this application scenario, we formulate the problem in a way different from the conventional next-item prediction.
%To fit for this application scenario, we formulate the problem as a next-$n$-day prediction, rather than the conventional next-item prediction.

To fit for this application scenario, we formulate both the IR and UT as a next-$n$-day prediction problem.
%\subsubsection{Interaction Matrix}
%In merchants' sales data, when a user $u$ purchased an item $i$ at time $t$, a record $(u,i,t)$ was logged.
When a user $u$ purchases an item $i$ at time $t$, a record $(u,i,t)$ is logged. 
%Merchants have accumulated lots of these records in their private channels, and want to utilize them for better marketing.
Given the raw logs $\{(u,i,t)\}$, the following data-processing method is applied for the next-$n$-day prediction: we create a dataset $\gD=\{(x_{u,t}, y_{u,t}):t\in\{1,2,...,T_u\}|u\in\{1,2,...,N\}\}$, where $x_{u,t}$ represents user $u$'s purchases prior to day $t$, \ie, a sequence of purchased items, and $y_{u,t}$ is any item purchased during the next $n$ days $[t, t+n)$, and $T_u=T$ is the number of days.
%which is the same for all users.

%there are usually two ways to process the data: one is to directly create a user-item interaction matrix $\mS_{ui}$ with the value $s_{u,i}$ counting the occurring times of $(u,i)$ pair in the logs \cite{su2009survey}, while the other one is to formulate the problem as a 

%In the conventional next-item prediction, the dataset $\gD$ is defined similarly, and $x_{u,t}=\{y_{u,1:(t-1)}\}$ represents $u$'s purchases prior to $t$-th purchase $y_{u,t}$ and $T_u$ is the number of purchases of $u$ \cite{covington2016deep}.

%In this paper ,li2019multi,yi2019sampling,yang2020mixed,zhou2021contrastive

\begin{figure}[thpb]
  \begin{center}
  \includegraphics[width=\columnwidth]{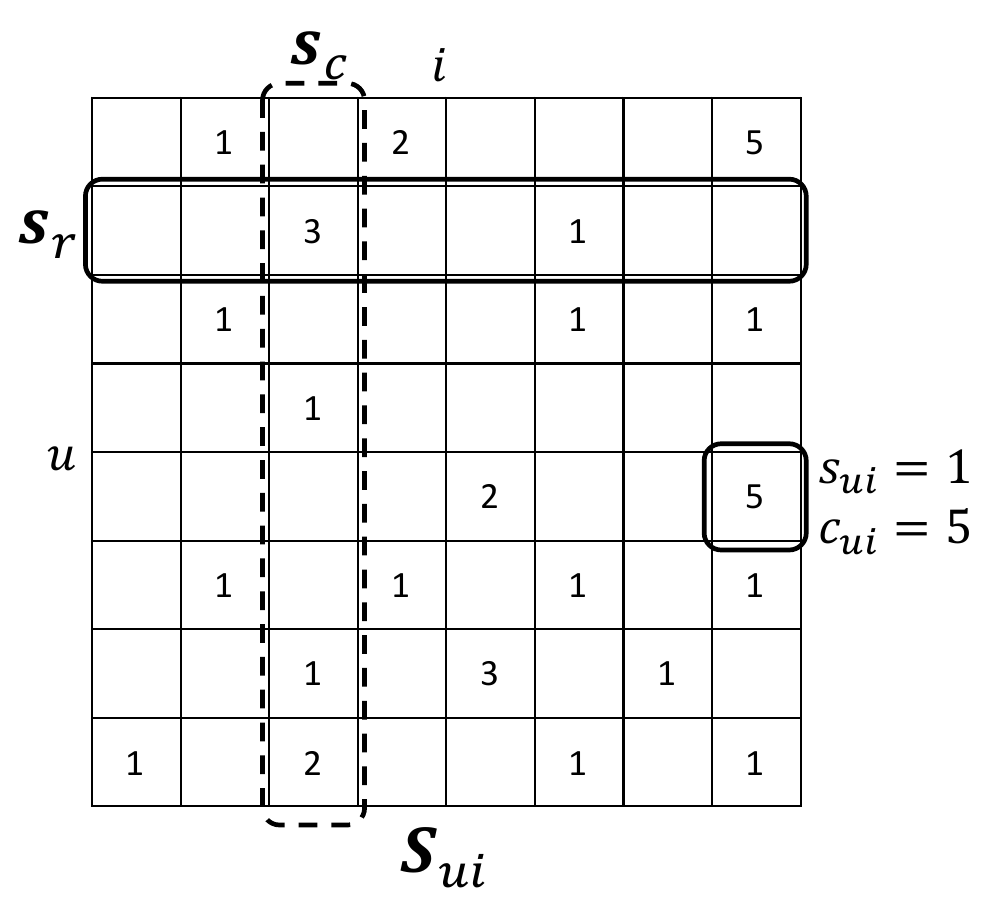}
  \end{center}
  \caption{An illustration of the user-item interaction matrix $\mS_{ui}$. We use $s \in \{0,1\}$ to denote whether there is an interaction between a user and an item, and $c \in \mathbb{N}$ to count the interaction times, \eg, $c_{ui}$ denotes the interaction times between $u$ and $i$. The row vector $\vs_r$ records the interactions between one user and all the items, and the column vector $\vs_c$ contains the interactions between one item and all the users. The ultimate goal is to solve the unknown entries in the matrix $\mS_{ui}$, but it is hard to model the whole matrix directly. So conventionally we either model the vectors $\vs_r$ or $\vs_c$ with the multinomial distribution or model the scalar $s$ with the Bernoulli distribution.}
  \label{fig:mo}
\end{figure}

The dataset $\gD$ forms a user-item interaction matrix $\mS_{ui}$, as shown in Fig.~\ref{fig:mo}, where the rows and columns represent $x_{u,t}$ and $y_{u,t}$, respectively.
%$y_{u,t}$ represents the item, and $x_{u,t}$ is actually the sequence of purchases. 
We call $x_{u,t}$ as the pseudo-user, and all possible sequences of purchases form the pseudo-user set. Without loss of generality, we use $u$ to represent the pseudo-user $x_{u,t}$, and $i$ to represent $y_{u,t}$ in the rest of the paper.

In the matrix $\mS_{ui}$, the entries are either the counts of $(u,i)$ interactions $c_{ui}$ or unknown. We have all the users form the set $\sU=\{u_1, u_2, ..., u_M\}$, and all the items form the set $\sI=\{i_1, i_2, ..., i_K\}$. In IR, given a user $u\in\sU$, we generate matched items from the item pool $\sI$. In UT, we shall find out the potential users out of all the users $\sU$ given an item $i\in\sI$. See Fig. \ref{fig:mo} for a detailed illustration.

\subsection{Modeling with the Bernoulli and Multinomial Distributions}
\label{sec:modeling}
%The two tasks of item recommendation and user targeting are both trying to estimate certain probabilities of the unknown entries in $\mS_{ui}$. 
Both the tasks of IR and UT try to estimate the value of the unknown entries in $\mS_{ui}$ with the probability. 
Traditionally, they are modeled with either the Bernoulli distribution on the scalar $s$ \cite{johnson2014logistic,he2017neural,rendle2020neural}, or the multinomial distribution on the vectors $\vs_r$ \cite{liang2018variational} or $\vs_c$, as depicted in Fig. \ref{fig:mo}. % and Tab. \ref{tab:distribution}.

% \begin{table}
% \centering
%   \caption{
%   The item Recommendation and user targeting can be modeled by either Bernoulli or Multinomial distributions. The detailed explanations are given in Sec. \ref{sec:modeling}.
%   }
%   \label{tab:distribution}
%   \begin{tabular}{lcc}
%     \toprule
%              & IR                                        & UT \\ \hline
%  Bernoulli   & $s \sim\rmB(\sigmoid(\phi_\theta(u,i)))$  & $s\sim\rmB(\sigmoid(\phi_\theta(u,i)))$  \\
%  Multinomial & $\vs_r \sim {\bf Mult}(N_u, \vp_u)$       & $\vs_c \sim {\bf Mult}(N_i, \vp_i)$  \\
%   \bottomrule
%   \end{tabular}
% \end{table}

%In Sec. xxx, we show that the two distributions can achieve the same results in modeling the two tasks.

\subsubsection{The Bernoulli Distribution}
\label{sec:bernoulli-d}
To conduct the IR and UT, we can model $s$ as a binary random variable drawn from a Bernoulli distribution. Then, we have $s\sim\rmB(\sigmoid(\phi_\theta(u,i)))$, which means $p(s=s_{u,i}|u,i)=\sigmoid(\phi_\theta(u,i))$, where $\sigmoid(\cdot)$ is the sigmoid function, $\phi_\theta(u,i)$ is the scalar output of the model parameterized by $\theta$, and it will be further illustrated in Sec. \ref{sec:arch}.

Conventionally, the training dataset is constructed with the positive and negative samples with $s\in\{0,1\}$. The likelihood of the dataset is the product of probabilities of all the single data point, \ie, $\prod_{u\in\sU, i\in \sI, (u,i) \in \gD_b}{p(s=s_{u,i}|u,i)}$. By maximizing the log-likelihood, we obtain the binary cross-entropy (BCE) loss:
\begin{equation}
\label{eq:ce_loss}
\begin{split}
l = - \frac{1}{|\gD_b|}\sum_{u\in\sU, i\in\sI, (u,i) \in \gD_b} 
    & s_{u,i}\log\sigmoid (\phi_{\theta}(u,i)) \\
    & + (1-s_{u,i})\log(1-\sigmoid (\phi_{\theta}(u,i))),
\end{split}
\end{equation}
where $s_{u,i} \in \{0,1\}$, and $s_{u,i}=1$ are the positive pairs in $\mS_{ui}$, and $s_{u,i}=0$ are the negative samples randomly sampled from $\mS_{ui}$ with the probability $p_n(u,i)$, and $\gD_b$ is the training dataset consisting of the positive and negative samples.

Many studies have employed the Bernoulli distribution for the IR and achieved state-of-the-art (SOTA) results \cite{he2017neural,rendle2020neural}. Theoretically, it should also work for the UT, and it will be further depicted in Sec. \ref{sec:optimum-b}.

\subsubsection{The Multinomial Distribution}
\label{sec:multinomial-d}
Within the multinomial distribution scope, the modeling objective can be either $\vs_r$ and $\vs_c$. Traditionally only $\vs_r$ is studied in the IR and UT research area. To the best of our knowledge, we do not find any research in the IR or UT that models $\vs_c$.

When modeling $\vs_r$, we assume that it is drawn from a multinomial distribution ${\bf Mult}(N_u, \vp_u)$ for a given user $u$. Here the total number of interactions $N_u = \sum_{i\in\sI} c_{u,i}$ of $u$, $\vp_u=(p_{u1},p_{u2},...,p_{uK})^T$ is a $K$-dimensional probability vector summing to 1 \cite{liang2018variational}. The $p_{uk} \coloneqq p(i=k|u)$ is the conditional probability modeled as
\begin{equation}
\label{eq:softmax}
  p_{uk} = \frac{\exp{\phi_\theta(u,k)}}{\sum_{j \in \sI}\exp{\phi_\theta(u,j)}},
\end{equation}
where $\phi_\theta(u,i)$ is the scalar measuring the similarity between $u$ and $i$, output of the model parameterized by $\theta$ as in Fig. \ref{fig:tr}.
%Although not explicitly discussed in many papers \cite{covington2016deep,yi2019sampling,li2019multi,cen2020controllable}, the multinomial distribution is the assumption of those research.
Although not explicitly discussed, many research works build upon the modeling with the multinomial distribution \cite{covington2016deep,yi2019sampling,li2019multi,cen2020controllable}.

When the modeling objective is $\vs_c$, it is assumed to follow the multinomial distribution ${\bf Mult}(N_i, \vp_i)$ for a given item $i$. Here the total number of interactions $N_i = \sum_{u\in\sU} c_{u,i}$ of $i$, $\vp_i=(p_{1i},p_{2i},...,p_{Mi})^T$ is a $M$-dimensional probability vector summing to 1, and $p_{mi} \coloneqq p(u=m|i)$ is the conditional probability.

% The UT seems to be the twin problem of IR, and its
By maximizing the multinomial loglikelihood \cite{liang2018variational}, we have the losses in Eqs. \ref{eq:u2i-loss} and \ref{eq:i2u-loss} for them respectively:
\begin{equation}
  \label{eq:u2i-loss}
    l = - \frac{1}{|\gD_m|}\sum_{u\in\sU, i\in\sI, (u,i) \in \gD_m} \log \frac{\exp{\phi_\theta(u,i)}}{\sum_{i' \in \sI}\exp{\phi_\theta(u,i')}},
\end{equation}

\begin{equation}
  \label{eq:i2u-loss}
    l = - \frac{1}{|\gD_m|}\sum_{u\in\sU, i\in\sI, (u,i) \in \gD_m} \log \frac{\exp{\phi_\theta(u,i)}}{\sum_{u' \in \sU}\exp{\phi_\theta(u',i)}},
\end{equation}
where $\gD_m$ is the training dataset consisting of only the positive user-item pairs.

% Modeling $\vs_r$ and $\vs_c$ are both trying to estimate certain probabilities of the unknown entries in $\mS_{ui}$, so we propose to model them jointly in our UniMatch Framework. This will be further illustrated in Sec. \ref{sec:loss}.

\section{A Unified User-Item Matching Framework}
\label{sec:method}
In this section, we first show that modeling with the Bernoulli and multinomial distributions are theoretically equivalent. We prove that in theory they can converge to the same optima by properly selecting the negative sampling methods for the Bernoulli distribution, and setting up the configurations for the multinomial distribution.

\begin{table}
\centering
  \caption{
  The BCE loss with different negative sampling probabilities $p_n(u,i)$ lead to different optima.
  }
  \label{tab:optima-bernoulli}
  \begin{tabular}{ll}
    \toprule
    $p_n(u, i) \propto$  & $\phi_\theta(u,i) \sim$ \\
    \midrule
    $\ptrain(u)$         & $\log \ptrain(i|u)$  \\
    $\ptrain(i)$         & $\log \ptrain(u|i)$  \\
    $\ptrain(u)\cdot\ptrain(i)$ & $\log\frac{\ptrain(u,i)}{\ptrain(u)\ptrain(i)}$\\
    1                    & $\log \ptrain(u,i)$  \\
  \bottomrule
\end{tabular}
\end{table}

Then, we elaborate on the proposed framework, \emph{UniMatch}, which consists of a bidirectional bias-corrected NCE loss, \emph{bbcNCE}, that models the $\vs_r$ and $\vs_c$ concurrently with the multinomial distribution leading the model convergence to $\ptrain(u,i)$, a two-tower architecture that can incorporate various models, 
%including the well-known Youtube-DNN/CNN/RNN/Transformers, etc, 
and an incremental training procedure that is tailored to the application of the private domain marketing.

The incremental training mechanism enables the model training to avoid learning from highly biased user-item distributions.
The bbcNCE loss allows us to train only one model and then infer one set of user and item embeddings, which can be used for both the IR and UT. We choose the bbcNCE loss originating from the multinomial distribution over the equivalent loss setup from the Bernoulli distribution, because we empirically unveil the discovery that the former produces better, more robust results, and saves training costs dramatically as in Sec. \ref{sec:exp}.

\begin{table*}
\centering
  \caption{
  The optima of the SSM loss and losses with different settings of Eq.~\ref{eq:j-infonce}. We propose to use the bbcNCE as the loss of our UniMatch framework for both item recommendation and user targeting.
  }
  \label{tab:optima-multinomial}
  \begin{tabular}{llll}
    \toprule
    Settings        & Objective          & $\phi_\theta(u,i) \sim$              & Loss      \\
    \midrule
    N/A             & $\vs_r$            & $\log \ptrain(i|u)$                  & SSM \cite{jean2015using}\\
    \midrule
    $\alpha=1, \delta_\alpha=\beta=\delta_\beta=0$   & $\vs_r$            & \multirow{2}{*}{$\log \frac{\ptrain(u,i)}{\ptrain(u)\ptrain(i)}$} & InfoNCE~\cite{oord2018representation} \\
    $\alpha=\beta=1$, $\delta_\alpha=\delta_\beta=0$ & $\vs_r$, $\vs_c$   &                                      & SimCLR~\cite{chen2020simple} \\
    \midrule                                     
    $\alpha=\delta_\alpha=1$, $\beta=\delta_\beta=0$ & $\vs_r$            & $\log \ptrain(i|u)$                  & row-bcNCE \\
    $\alpha=\delta_\alpha=0$, $\beta=\delta_\beta=1$ & $\vs_c$            & $\log \ptrain(u|i)$                  & col-bcNCE \\
    $\alpha=\delta_\alpha=\beta=\delta_\beta=1$      & $\vs_r$, $\vs_c$   & $\log \ptrain(u,i)$                  & \textbf{\textit{bbc}NCE} \\
  \bottomrule
\end{tabular}
\end{table*}

\subsection{The equivalence between modeling with the Bernoulli and multinomial distributions}
\label{sec:connection-d}
When modeling with the Bernoulli distribution, we deal with $p(s|u,i)$ directly, while we study $p(i|u)$ or $p(u|i)$ with the multinomial distribution. In order to bridge the gap between modeling with these two distributions, we uncover the theoretical connection between these two modeling strategies, and prove that they can converge to the same optima.

\subsubsection{Optima of modeling with the Bernoulli distribution}
\label{sec:optimum-b}
We derive the optima of modeling with the Bernoulli distribution for various negative sampling methods. Inspired by Noise Contrastive Estimation (NCE) \cite{gutmann2010noise}, we assume the positive samples of the training dataset form the set $\sX$, and the negative samples form the set $\sY$. The negative samples are randomly sampled based on a certain distribution $p_n(u,i)$. %$p_n(\cdot)$

Assume $\sX=\{\vx_1, \vx_2, ..., \vx_L\}$ contains $L$ samples, and $\sY=\{\vy_1, \vy_2, ..., \vy_F\}$ contains $F$ samples, and $\sZ=\sX \cup \sY = \{\vz_1, \vz_2, ..., \vz_{L+F}\}$ contains all the $L+F$ samples. Here $\vx_l \coloneqq (u,i)$, $\vy_f \coloneqq (u,i)$ and $\vz_j \coloneqq (u,i)$, where $u\in\sU$ and $i\in\sI$. We assign each sample $\vz_{j}$ a binary class $C_{j}$: $C_{j}=1$ if $\vz_{j}$ comes from $\sX$, and $C_{j}=0$ if $\vz_{j}$ is from $\sY$.

% the observed matrix $\mS_{ui}$ is generated from a probability distribution $p(\vz)$, where $\vz \in \{(u,i)\}$. We model $p(\vz)$ by reformulating it as the binary classification problem, and thus define the class-conditional probabilities:
We assume the joint probability of the positive samples in $\sX$ is parameterized by $\Tilde{\theta}$ as $\pmodel(u,i;\Tilde{\theta})=\pmodel(\vz;\Tilde{\theta})$. So we have the conditional probabilities:
\begin{equation}
  \label{eq:nce_bi}
    p(\vz|C=1;\Tilde{\theta}) = \pmodel(\vz;\Tilde{\theta}) \qquad p(\vz|C=0;\Tilde{\theta}) = p_n(\vz).
\end{equation}

The posterior probabilities are:
\begin{equation}
\label{eq:nce_pos}
\begin{split}
  p(C=1|\vz;\Tilde{\theta}) & = \frac{\pmodel(\vz;\Tilde{\theta}) P(C=1)}{\pmodel(\vz;\Tilde{\theta}) P(C=1) + p_n(\vz) P(C=0)} \\
        & = \frac{1}{1+\exp(-G(\vz;\Tilde{\theta}))} = \sigmoid(G(\vz;\Tilde{\theta})),
\end{split}
\end{equation}
where
\begin{equation}
\label{eq:nce_g}
G(\vz;\Tilde{\theta}) = \log\frac{\pmodel(\vz;\Tilde{\theta}) P(C=1)}{p_n(\vz) P(C=0)},
\end{equation}
and we also have
$$
p(C=0|\vz;\Tilde{\theta}) = \frac{1}{1+\exp(G(\vz;\Tilde{\theta}))}.
$$
So the log-likelihood is:
\begin{equation*}
\begin{split}
  l(\Tilde{\theta}) = \sum^{L+F}_{j=1} & C_j \log P(C_j=1|\vz_j;\Tilde{\theta}) \\
    & + (1-C_j) \log P(C_j=0|\vz_j;\Tilde{\theta}).
\end{split}
\end{equation*}

Through optimizing the binary classification using the samples in $\sZ$ and the corresponding binary classes $C$, we are actually recovering the modeling of $s$ with the Bernoulli distribution as in Eq. \ref{eq:ce_loss}. The dot product $\phi_{\theta}(u,i)$ of Eq. \ref{eq:ce_loss} is $G(\vz;\theta)$ in Eq. \ref{eq:nce_g}:
\begin{equation}
%   \label{eq:nce_opt}
    \phi_{\theta}(u,i) = \log\frac{\pmodel(\vz;\Tilde{\theta}) P(C=1)}{p_n(\vz) P(C=0)}.
\end{equation}

It is shown that $\pmodel(\vz;\Tilde{\theta})$ will converge to the empirical distribution $\ptrain(\vz)$ in \cite{gutmann2010noise}. As $\vz \coloneqq (u,i)$, we will have the following equation: 
\begin{equation}
\label{eq:nce_opt}
\phi_{\theta}(u,i) \approx \log \frac{\ptrain(u,i)}{p_n(u,i)} + C',
\end{equation}
where $C'$ denotes some constant.

Different negative sampling strategies $p_n(u,i)$ will lead to very different optimal $\phi_{\theta}(u,i)$. For example, if we randomly sample $\tilde{n}$ items for each positive $(u,i)$ pair to form the negative samples with the user $u$, then we have $p_n(u,i)=\ptrain(u)\cdot 1/K$, where $\ptrain(u)$ is the empirical marginal probability of the $u$. Substitute $p_n(u,i)$ into Eq. \ref{eq:nce_opt}, we have $\phi_{\theta}(u,i) \approx \ptrain(i|u)$. Similarly, we can derive the results in Tab. \ref{tab:optima-bernoulli} for other negative sampling methods. Specifically, when sampling with the uniform probability, we would have $\phi_\theta(u,i) \sim \log \ptrain(u,i)$, which could be used for both the IR and UT.

\subsubsection{Optima of modeling with the multinomial distribution}
\label{sec:optimum-m}
When modeling $\vs_r$ or $\vs_c$ with the multinomial distribution, we have the losses in Eqs. \ref{eq:u2i-loss} and \ref{eq:i2u-loss}. The vocabularies of the user set $\sU$ and item set $\sI$ are very large, so the calculation of the partition functions in the losses are very time-consuming and memory-exhausting, causing problems during the optimization \cite{jean2015using}. In practice, it can be solved by implementing with the sampled softmax loss (SSM) \cite{jean2015using,covington2016deep}. The InfoNCE loss \cite{oord2018representation} provides an alternative implementation with the sampling bias attached as in CLRec \cite{zhou2021contrastive}.

In our applications of the IR and UT, we propose to model $\vs_r$ and $\vs_c$ concurrently by combining the two losses into one, and implement it with the bias-corrected NCE loss inspired by the InfoNCE loss. The resulting loss is shown in Eq. \ref{eq:j-infonce}:

%\begingroup\makeatletter\def\f@size{8}\check@mathfonts
%\begin{equation}
\begin{multline}
\label{eq:j-infonce}
        l = -\frac{1}{|\mS_{u,i}|}\sum_{u\in\sU,i\in\sI, s_{u,i}=1}
         \alpha \cdot \log \frac{h(u,i)}{h(u,i) + \sum_{i'\in\sI_u}h(u,i') } \\
         + \beta \cdot \log \frac{o(u,i)}{o(u,i) + \sum_{u'\in\sU_i} o(u,i') }
,
\end{multline}
%\end{equation}
%\endgroup
where $h(u,i)=\exp (\phi_{\theta}(u,i) - \delta_\alpha \log \ptrain(i))$, and $o(u,i)=\exp (\phi_{\theta}(u,i) - \delta_\beta \log \ptrain(u))$, $\sI_u \subset \sI$ and $\sU_i \subset \sU$ contain hundreds or thousands of in-batch negative samples as in Tab. \ref{tab:infonce-loss}, and $\ptrain(i)$ and $\ptrain(u)$ are empirical marginal distributions calculated using the training data. $\alpha$, $\beta$, $\delta_\alpha$ and $\delta_\beta$ are binary numbers. $\delta_\alpha \log \ptrain(i)$ and $\delta_\beta \log \ptrain(u)$ are the bias correction terms, which `correct' the biases caused by the in-batch negative sampling.

\begin{figure*}[thpb]
  \begin{center}
  \includegraphics[width=0.7\textwidth]{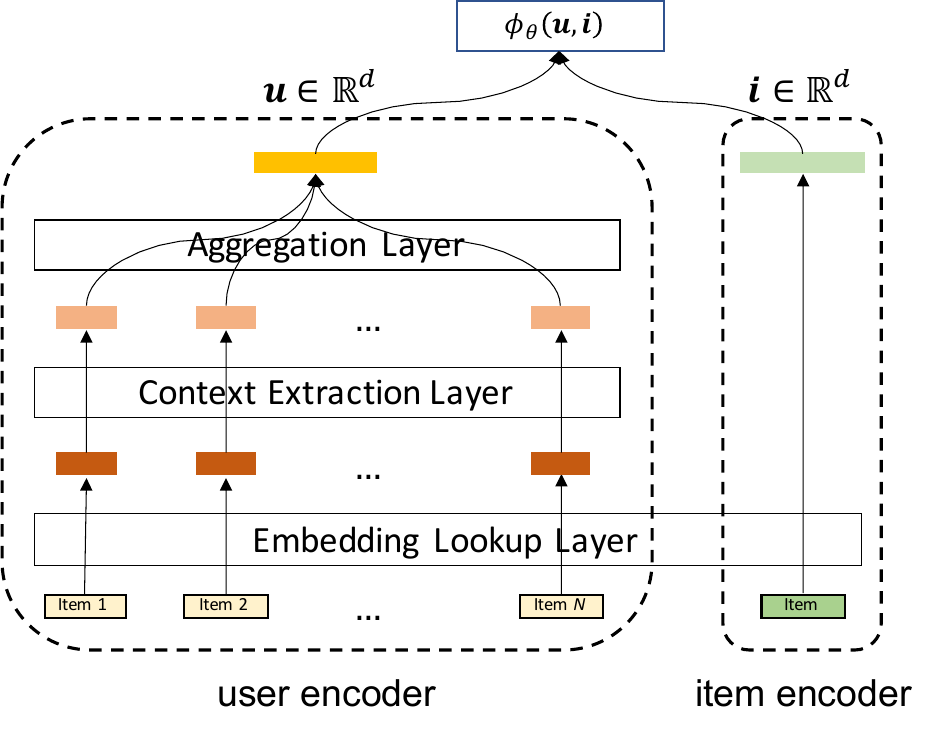}
  \end{center}
  \caption{The model architecture of the \textit{UniMatch} framework. Users' behavior sequences and items' features go through the encoders separately, and output the $d$-dimensional representation vectors $\vu$ and $\vi$. The two encoders share the same item embedding lookup table. Their $l2$-normalized dot product is rescaled by the temperature hyperparameter $\tau$ to obtain $\phi_\theta(u,i)$ as in Eq. \ref{eq:dot}, which is then passed to the loss functions, \eg, Eq. \ref{eq:ce_loss} and \ref{eq:j-infonce}.
  }
  \label{fig:tr}
\end{figure*}

We call the first part row loss and the second part column loss (See Fig.~\ref{fig:mo}). As shown in \cite{oord2018representation}, the row loss can be regarded as an approximation of the loss in Eq. \ref{eq:u2i-loss}, and the column loss as an approximation of the loss in Eq. \ref{eq:i2u-loss}.

The InfoNCE and SimCLR losses are the special cases of Eq. \ref{eq:j-infonce} when the bias correction terms are omitted, \ie, $\delta_\alpha=\delta_\beta=0$. Then we have the InfoNCE loss by setting $\alpha=1, \beta=0$, and the SimCLR loss with $\alpha=1 = \beta=0$ as in Tab. \ref{tab:optima-multinomial}.

Different settings will lead to different optima of $\phi_\theta(u,i)$. It has been shown that the setting for InfoNCE will have the optimum $\exp(\phi_\theta(u,i)) \propto \frac{\ptrain(i|u)}{\ptrain(i)}$ in \cite{oord2018representation}. It is straightforward that the SimCLR loss has the same optimum.

In the cases that the bias terms are retained, \ie, $\delta_\alpha=\delta_\beta=1$, the optimum of $\phi_\theta(u,i)$ would be different. 
%We name the resulting loss \textbf{bcNCE}, short for bias-corrected NCE.
It results in an NCE loss with the bias-correction.
Different from the InfoNCE loss that converges at the point where the mutual information between users and items is maximized, the bias-corrected NCE losses (bcNCE) converge when the conditional probability is fitted by $\phi_\theta(u,i)$. For example, when $\alpha=1, \beta=0$, the loss degenerates to the row loss in Eq. \ref{eq:j-infonce}, and the optimum is $\frac{\exp(\phi_\theta(u,i))}{\ptrain(i)} \propto \frac{\ptrain(i|u)}{\ptrain(i)}$. Then we have $\phi_\theta(u,i) \propto \log \ptrain(i|u)$. Analogously, we have $\phi_\theta(u,i) \propto \log\ptrain(u|i)$ for the setting $\alpha=0, \beta=1$ as in Tab. \ref{tab:optima-multinomial}.

When $\alpha=\beta=\delta_\alpha=\delta_\beta=1$, the optimum is not trivial to derive. 
%We show our proof in the following.
The proof is given as follows:

In this setting, we have $\phi_\theta(u,i) \propto \log \ptrain(i|u)$ for the row loss of Eq. \ref{eq:j-infonce} at its optimum, so we can assume 
% \begin{multline}
\begin{equation}
\label{eq:phi_row}
    \begin{split}
\phi_\theta(u,i) & = \log\ptrain(i|u) + f(u) \\
    & = \log\ptrain(u,i) - \log\ptrain(u) + f(u),
    \end{split}
% \end{multline}
\end{equation}
for a given $u$ and any $i\in \sI$, where $f(\cdot)$ is an arbitrary function depending on $u$ only. Similarly, in the optimum of the column loss, we have 
\begin{equation}
\label{eq:phi_col}
    \phi_\theta(u,i) = \log\ptrain(u,i) - \log\ptrain(i) + g(i)
\end{equation}
for a given $i$ and any $u\in \sU$, and $g(\cdot)$ is an arbitrary function depending on $i$ only.

% In item recommendation and user targeting, we propose to retain the bias correction terms and name the resulting loss \textit{bc}NCE, short for bias-corrected NCE. Specifically, in Tab. \ref{tab:optima-multinomial}, \textit{row/col}-bcNCE stands for row-wise/cloum-wise-bias-corrected NCE, and bbcNCE represents bidirectional-bias-corrected NCE.

% In this work, we propose to use the setting $\alpha=\delta_\alpha=\beta=\delta_\beta=1$, \ie, \textit{bbc}-NCE to do item recommendation and user targeting simultaneously. In the following of this subsection, we prove that $\phi_\theta(u,i)$ converges to $\log \ptrain(u,i)$ in this setting. $\ptrain(u,i)$ is the joint distribution of $u$ and $i$ in the training data.

% As proved in \cite{oord2018representation}, the model trained with the InfoNCE loss will have the optimum $\exp(\phi_\theta(u,i)) \propto \frac{\ptrain(i|u)}{\ptrain(i)}$. 
% Following the above proof, we have the optimum of the row loss of Eq. \ref{eq:j-infonce} as $\frac{\exp(\phi_\theta(u,i))}{\ptrain(i)} \propto \frac{\ptrain(i|u)}{\ptrain(i)}$, so we can assume 

%$$
%\phi_\theta(u,i) = \log\ptrain(i|u) + f(u) = \log\ptrain(u,i) - %\log\ptrain(u)+ f(u)
%$$

Thus, from the equivalence of Eq. \ref{eq:phi_row} and \ref{eq:phi_col}, we have the equation that always holds for any $u$ and $i$: $- \log\ptrain(u) + f(u) \equiv - \log\ptrain(i) + g(i)$, so it must be some constant. Then we have $\phi_\theta(u,i) = \log\ptrain(u,i) + C'$, where $C'$ is some constant that is independent of $u$ and $i$. When $\phi_\theta(u,i)$ converges to $\log\ptrain(u,i)$, both parts in Eq.~\ref{eq:j-infonce} can reach their optima, so it is at least one of the solutions of the whole loss. We name the resulting loss \textbf{bbcNCE}, short for bidirectional-bias-corrected NCE. It is employed in our framework for the IR and UT, as further illustrated in Sec. \ref{sec:unimatch}.

%Various settings of $\alpha$, $\delta_\alpha$, $\beta$ and $\delta_\beta$ and their resulting optima are listed in Tab.~\ref{tab:optima-multinomial}.

As proved in the above sections, the optima of different settings of modeling with the Bernoulli and multinomial distributions are listed in Tab. \ref{tab:optima-bernoulli} and \ref{tab:optima-multinomial}. We can see that they can guide the parameterized models to converge at the same optima. Therefore, we can conclude that they are equivalent on modeling the user-item interactions depicted in Fig. \ref{fig:mo}.

\begin{table*}[!b]
\caption{The statistics of the experimental datasets, including two open datasets Amazon books and electronics as well as two real-world datasets from our QA clients.}
\label{tab:exp_stats}
% \begin{center}
% \begin{tabular}{c|ccc}
% \toprule
%       & \# records & \# items & \# users \\
% \midrule
% train & 990296 & 3076 & 6040 \\
% valid & 5125    & 1257   & 2698  \\
% test  & 4788    & 1257   & 2547  \\
% % all   & 9913    & 2514   & 3691 \\
% \bottomrule
% \end{tabular}
% \end{center}
\centering
\begin{tabular}{p{0.1\linewidth} | p{0.12\linewidth} p{0.12\linewidth} p{0.15\linewidth} p{0.1\linewidth} p{0.12\linewidth} p{0.12\linewidth} p{0.12\linewidth}}%{l|llllll}
\toprule
Data               & \# users  & \# items & \# interaction & time-span & avg. \#actions/user & avg. \#actions/item \\ 
\midrule
Books              & 536,409   & 338,739  & 6,132,506      & 31 months   & 11.4               & 18.1               \\
Electronics        & 3,142,438 & 382,246  & 5,566,859      & 31 months   & 1.8                & 14.6               \\
QA e\_comp         & 237,052   & 15,168   & 1,350,566      & 47 months   & 5.7                & 89.0               \\
QA w\_comp         & 867,107   & 507      & 2,762,870      & 24 months   & 3.2                & 5449.4             \\ 
\bottomrule
\end{tabular}
\end{table*}

\subsection{The UniMatch Framework}
\label{sec:unimatch}
We propose to model the IR and UT jointly in the \textit{UniMatch} framework. In details, the \textit{UniMatch} consists of the classical two-tower architecture, the bbcNCE loss proposed in the previous section and the incremental training procedure.

\subsubsection{Architecture}
\label{sec:arch}
We choose this architecture for two reasons. The first is that the users and items can be processed equivalently, while the other one is that there is no feature crossing occurring before the final logits $\phi_\theta(u,i)$, as shown in Fig.~\ref{fig:tr}. As a result, users' and items' embeddings can be inferred separately, and then the approximate nearest neighbor (ANN) search algorithm can be applied during serving \cite{liu2004investigation}.

The output of two towers are $d$-dimensional vectors $\vu=f_{\theta}(u) \in\sR^{d}$ and $\vi=g_{\theta}(i)\in\sR^{d}$, where $\theta$ is the model parameter. The dot product $\langle \vu | \vi \rangle$, or the function of it is used as the sufficient statistics of the probability distributions defined in Sec.~\ref{sec:connection-d}. We find that l2-normalizing $\vu$ and $\vi$ and then rescaling the dot product by the temperature $\tau$ lead to better and robust results:
\begin{equation}
\label{eq:dot}
\phi_{\theta}(u,i) = \frac{1}{\tau} \frac{\langle \vu | \vi \rangle}{||\vu||_2 ||\vi||_2}.
\end{equation}

\begin{itemize}
\item \textbf{User Encoder}. In this work, users' behavior sequences are used as the features of the user encoder, so any sequential model can be adopted here. 
For example, the CNN \cite{lecun1990,alexnet2012}, RNN (GRU \cite{cho2014learning}, LSTM \cite{gers2000learning}), Transformer \cite{vaswani2017} and their enormous variants can be used here. 
In fact, they have been widely applied in the item recommendation, such as Caser \cite{tang2018personalized}, GRU4Rec \cite{hidasi2015session}
SASRec \cite{kang2018self}, and etc.

We abstract the user encoder into 3 parts, embedding lookup layer, context extraction layer and aggregation layer. Through the lookup layer, item-ids are turned into vectors. In the context extraction layer, we extract and fuse the contextual information for each item in the behavior sequence with CNN/RNN/Transformer. Finally, we aggregate all the sequential item embeddings with max/mean/last/attention pooling methods. Here the last pooling means picking the last embedding in the sequence, and the attention pooling means summing over all the embeddings with learned attention weights. 
%[Put some equations here?] 

\item \textbf{Item Encoder}. The item encoder takes item features as the input and outputs a representative vector. In this work, we obtain the item vectors directly from the lookup table. 
\end{itemize}

Our framework is model agnostic, and in case that other formats of data is taken as the input, the corresponding models can be used to replace the sequential models.

%Since our framework is model agnostic, we can train other formats of input data other than sequential data can be trained with the framework as well. The flexible design of our framework ensures that 

\subsubsection{Loss Function}
\label{sec:loss}
As illustrated in Sec. \ref{sec:preliminaries}, we can choose to model $\mS_{ui}$ with either the Bernoulli or multinomial distributions to do the IR and UT simultaneously. And we have proved that they could lead to the same optima in Sec. \ref{sec:connection-d}. This implies that we can employ either the BCE loss in Eq. \ref{eq:ce_loss} or the NCE loss in Eq. \ref{eq:j-infonce} in our framework. 

Both the BCE loss with the uniform negative sampling and the bbcNCE converge at the joint probability $\ptrain(u,i)$. Theoretically, the two losses with the corresponding settings should perform well in both IR and UT. Our experiments show that the bbcNCE loss yields better and more robust results across all 4 datasets. In addition, bbcNCE requires $1/10 \sim 1/5$ of the training time of the BCE, thus reduce the cost very much. So we choose the bbcNCE as the loss of our \textit{UniMatch} framework.

\subsubsection{Incremental training}
Incremental training feeds the training data sequentially based on the absolute time. It has two advantages compare to feeding all the training data randomly instead. First, in this setting, we train the model every month from the saved checkpoint using the latest 1-month training data. This will save lots of cost compare to training with all the data in the past dozens of months that are shuffled randomly. The other is that the results are much better. When trained with the latest 1-month data, the model parameters will shift to fit the updated user-item distribution, and thus boost the results on predicting the near future.

\section{Experiments}
\label{sec:exp}
We verify whether the proposed framework is able to yield the SOTA results for the IR and UT tasks on two public datasets and two real-world datasets.
Comprehensive experiments are designed to compare the two modeling strategies of the Bernoulli and multinomial distributions, and different losses within the multinomial distribution scope are evaluated. In addition, we show that our \textit{UniMatch} framework is model agnostic. It can adopt different models and produce better results consistently. Finally, we show that the incremental training is necessary for our applications.

\begin{table}
\centering
  \caption{
  The training data samples of the losses derived from modeling with multinomial distributions, \eg, SSM, InfoNCE, bbcNCE and etc. 
  %The `item\_seq' column records the user's historical behavior sequence as the input of the user encoder in Fig. \ref{fig:tr}. The `item\_id' column is the id feature of the item encoder. The last two columns are the bias-correction terms calculated from the marginal distributions of users and items in the training data, as illustrated in Eq. \ref{eq:j-infonce}. They are used in the bcNCE losses. Each record itself is the positive ui pair, and the in-batch negative sampling use the users or items in the same batch to form the negative ui pairs. For example, assume the following ui pairs are in the same batch during training, then for the record $(406690, 19273)$, item $19273$ is the positive item of user $406690$, and $\sI_u=\{39415,21632,176520,272267\}$ are the corresponding item set of negatives as in Eq. \ref{eq:j-infonce}.
  }
  \label{tab:infonce-loss} 
  \begin{tabular}{lllll}
  \toprule
    user\_id & item\_seq                    & item\_id & $\log(p(u))$ & $\log(p(i))$ \\
  \midrule
    406690   & 27886 755 4609 1319          & 19273    & -11.83447 & -9.34957  \\
    357729   & 8926 42571 9499              & 39415    & -9.63725  & -10.91818 \\
    392972   & 14172 6887 177888            & 21632    & -11.42901 & -11.83447 \\
    354500   & 85014 850  16291             & 176520   & -11.42901 & -10.73586 \\
    15839    & 10528 690 173  17            & 272267   & -11.42901 & -12.52762 \\
  \bottomrule
  \end{tabular}
\end{table}

\begin{table}
\centering
  \caption{
  The training data samples of the BCE loss derived from modeling with Bernoulli distributions. 
  %The `item\_seq' column records the user's historical behavior sequence as the input of the user encoder in Fig. \ref{fig:tr}. The `item\_id' column is the id feature of the item encoder. Records with `label' 1 are positive samples, and `label' 0 are negative samples that are sampled with certain distributions $p_n(u,i)$ (See Tab. \ref{tab:optima-bernoulli}). The ratio between positive and negative samples is $1:1$.
  }
  \label{tab:ce-loss} 
  \begin{tabular}{llll}
  \toprule
    user\_id & item\_seq                         & item\_id & label \\
  \midrule
    406690   & 27886 755 4609 1319               & 19273    & 1  \\
    357729   & 8926 42571 9499                   & 39415    & 1  \\
    394560   & 60076 5568 186 11 7 274408        & 16751    & 0  \\
    392972   & 14172 6887 177888                 & 21632    & 1  \\
    391953   & 70 20167 171                      & 6493     & 0     \\
  \bottomrule
  \end{tabular}
\end{table}

\begin{table*}[t]
\caption{The statistics of the 4 experimental datasets after splitting into train and test. }
\label{tab:train_test}
\centering
\begin{tabular}{ll|ll|ll}
\toprule
   &             & Amazon Books & Amazon Electronics & QA e\_comp & QA w\_comp \\
\midrule
   & train data  & 2,985,163      & 451,283             & 504,500  & 328,770  \\
\midrule
IR & \# test users  & 43,867        & 7,916               & 4,685    & 29,168   \\
   & \# item pool   & 67,967        & 14,118              & 1,943    & 221     \\
   & \# top-$n$ items & 10           & 10                 & 10      & 5       \\
   & \# negatives   & 99           & 99                 & 99      & 49      \\
\midrule
UT & \# test items  & 27,541        & 4,708               & 1,324    & 203     \\
   & \# user pool   & 317,667       & 207,060             & 30,439   & 171,354  \\
   & \# top-$n$ users & 10           & 10                 & 10      & 5       \\
   & \# negatives   & 99           & 99                 & 99      & 49     \\
\bottomrule
\end{tabular}
\end{table*}

\subsection{Experimental Setup}

\subsubsection{Datasets}
\label{sec:ds}
We use two public datasets, the Amazon books and electronics data\footnote{\url{http://jmcauley.ucsd.edu/data/amazon/index.html}} as well as two real-world datasets collected from two QuickAudience clients. The statistics of the data is shown in Tab.~\ref{tab:exp_stats}.
\begin{itemize}
\item \textbf{Amazon}. We choose two commonly used datasets, Amazon books and electronics from \textit{Amazon.com}. The datasets span from May 1996 to July 2014. We utilize the data from January 2012 to July 2014 in our experiments. For Amazon books and electronics, each sample's behavior sequence is truncated at the length of 20 and 36, respectively.
\item \textbf{QuickAudience clients}. We use e\_comp and w\_comp to represent the two merchant clients, and these two datasets span 47 and 24 months. 
Compared to Amazon datasets, they have comparable number of users and interactions, but the number of items are significantly less, so they are less sparse as in Tab.~\ref{tab:exp_stats}. We truncate the samples at the length of 29 and 18 for e\_comp and w\_comp, respectively. 
\end{itemize}

In our experiments, the next-$n$-day prediction is set to predict for the next month. With the whole dataset spanning $T$ months, we split the the data into train, validation and test data as $(0, T-1]$, $(T-2, T-1]$ and $(T-1, T]$.

In the train/validation/test data, we filter out the users/items who interact with less than 3 items/users. To train the model, we adopt the incremental training method, and consume the data consecutively according to the date $t$. In other words, we feed data of $t=1$ first and train for some epochs, and then followed by $t=2,3,..,T-1$. 
%Compared to the conventional way of randomly shuffling all the data and then feeding into training, the incremental training yield much better results.

%For the models trained with the losses derived from Multinomial and Bernoulli distributions, the training data is different. For example, the bcNCE losses shall have the pre-calculated bias terms as the input. Explicit negative samples must be prepared for the BCE loss, and the sampling ratio is $1:1$ of positive samples. So the training data is doubled compare to the bcNCE loss.
The losses derived from the multinomial and Bernoulli distributions require different input data formats. 
The differences are shown in Tab. \ref{tab:infonce-loss} and \ref{tab:ce-loss} with data samples.
To be more specific, the losses like bbcNCE require the bias-correction terms pre-calculated from the empirical distributions of users and items in the training data, as illustrated in Eq. \ref{eq:j-infonce}.
Each record is the positive user-item pair, and the in-batch negative sampling use the users or items in the same batch to form the negative user-item pairs. 
%For example, assume the following ui pairs are in the same batch during training, then for the record $(406690, 19273)$, item $19273$ is the positive item of user $406690$, and $\sI_u=\{39415,21632,176520,272267\}$ are the corresponding item set of negatives as in Eq. \ref{eq:j-infonce}.
On the other hand, for the BCE loss derived from modeling with the Bernoulli distribution, the records with label 1 are positive samples, and label 0 are negative samples that are sampled with certain distributions $p_n(u,i)$ as in Tab. \ref{tab:optima-bernoulli}. 
The ratio between positive and negative samples is $1:1$.

\begin{table*}[]
\caption{The hyperparameters of all the datasets modeled with Bernoulli or the multinomial distributions.}
\label{tab:hyper-param}
\centering
\begin{tabular}{l|ll|ll|ll|ll}
\toprule
                & \multicolumn{2}{c|}{Amazon books} & \multicolumn{2}{c|}{Amazon Electronics} & \multicolumn{2}{c|}{QA e\_comp} & \multicolumn{2}{c}{QA w\_comp} \\
Hyperparameters & Bernoulli      & Multinomial     & Bernoulli         & Multinomial        & Bernoulli   & Multinomial   & Bernoulli   & Multinomial   \\
\midrule
Batch-size      & 128            & 64              & 256               & 64                 & 128         & 64            & 128         & 64            \\
Temperature     & 0.1667         & 0.1667          & 0.5               & 0.5                & 0.25        & 0.125         & 0.125       & 0.1           \\
Epochs          & 8              & 3               & 6                 & 2                  & 6           & 2             & 10          & 2             \\
\bottomrule
\end{tabular}
\end{table*}

The test data of the IR and UT are prepared separately. 
The statistics of the 4 experimental datasets after splitting to train and test are shown in Tab. \ref{tab:train_test}.
%(See Tab. \ref{tab:train_test}). %They are the same for all the models and losses. 
We describe the table using the Amazon Books dataset as an example. 
Its train data contains 2,985,163 records as the positive samples. 
For the IR, the number of test users is 43,867. 
Each user has 1 positive item and 99 negative items that is sampled randomly from the item pool of 67967 items in total. 
Our experimental models predict top 10 items out of the 100 candidates for each user, and the results are evaluated using the metrics Recall@$10$ and NDCG@$10$ depicted in the following section. The same logic applies in the UT.
%In item recommendation, for each candidate user, we randomly pick 1 item as the positive from their interacted items in $(T-1, T]$, and randomly pick 99 items as the negatives from the non-interacted items. In user targeting, we apply the same logic for each candidate item.

\subsubsection{Hyperparameters}
We have three hyperparameters to be tuned in the experiments, \ie, temperature $\tau$, batch-size and the number of epochs, and choose the hyperparameters based on the validation data through grid search.
Different datasets modeled with different distributions have their own specific hyperparameters, and the grid search results are listed in Tab. \ref{tab:hyper-param}. The dimension $d=16$ is adopted for all the datasets in the experiments.

\subsubsection{Evaluation Metrics}
%We employ two commonly used metrics to evaluate the top-$N$ ranked items/users in IR and UT, \ie, Recall@$N$ and NDCG@$N$. NDCG stands for Normalized Discounted Cumulative Gain. 
We employ two commonly used metrics, Recall and Normalized Discounted Cumulative Gain (NDCG), and report Recall@$N$ and NDCG@$N$ to evaluate the top-$N$ ranked items/users in the IR and UT.
Another popular metric HitRate@$N$ is the same as Recall@$N$ when there is only 1 positive in the candidate pool, so we will not repeat its results here.

For the IR, the two metrics %Recall@$N$ and NDCG@$N$ of top-$N$ ranked items 
are defined as follows: 
% \begin{equation}
% \label{eq:hr}
% \text{HitRate@}N = \frac{1}{|\sU|}\sum_{u\in\sU} \delta(|\hat{\gI}_{u,N}\cap \gI_u|>0),
% \end{equation}
\begin{equation}
\label{eq:recall}
\text{Recall@}N = \frac{1}{|\sU|}\sum_{u\in\sU} \frac{|\hat{\gI}_{u,N}\cap \gI_u|}{\text{Min}(|\gI_u|, N)},
\end{equation}

\begin{equation}
\label{eq:ndcg}
\text{NDCG@}N = \frac{1}{|\sU|}\sum_{u\in\sU} \frac{1}{Z_u} \sum^{N}_{n=1} \frac{\delta(\hat{i}_{u,n} \in \gI_u)}{\log_2(n+1)},
\end{equation}
where $\hat{\gI}_{u,N}$ denotes the top-$N$ ranked items for $u$, $\hat{i}_{u,n}$ is $n$-th recommended item, $\gI_u$ is the set of ground-truth items, and $\delta(\cdot)$ is the indicator function. $Z_u$ is the normalization constant denoting the best possible discounted cumulative gain (DCG) for the user $u$, which means that all the ground-truth items are ranked at the top. For the UT the metrics are defined symmetrically.

%For UT, the Recall@$N$ and NDCG@$N$ of the top-$N$ targeted users of items are defined similarly, \ie, 
%$$\text{Recall@}N = \frac{1}{|\sI|}\sum_{i\in\sI} \frac{|\hat{\gU}_{i,N}\cap \gU_i|}{\text{Min}(|\gU_i|,N)},$$ 
%and 
%$$\text{NDCG@}N = \frac{1}{|\sI|} \sum_{i\in\sI} \frac{1}{Z_i} \sum^{N}_{n=1} \frac{\delta(\hat{u}_{i,n} \in \gU_i)}{\log_2(n+1)},$$ 
%where $\hat{\gU}_{i,N}$ denotes the set of top-$N$ picked users for $i$, $\hat{u}_{i,n}$ is the $n$-th targeted user, $\gU_i$ is the set of users who actually purchase $i$, $Z_i$ is the normalization constant denoting the best possible DCG for the item $i$.

In the experiments, we use Recall/NDCG@$10$ for the IR and UT across all the datasets except for QA w\_comp. We measure QA w\_comp with Recall/NDCG@$5$ due to its small number of items as in Tab. \ref{tab:exp_stats}. The NDCG measures the ranking status of the recommended items or targeted users, and contains more subtle information of the results, 
therefore we use it to select the hyperparameters as well.
%so we use NDCG as our main metric. For example, we use NDCG to select the hyperparameters, and always report NDCG in the experiments.

% $\text{HitRate@}N = \frac{1}{|\sI|}\sum_{i\in\sI} \delta(|\hat{\gU}_{i,N}\cap \gU_i|>0)$ and
% \begin{equation}
% \label{eq:ut_precision}
% \text{Precision@}N = \frac{1}{|\sI|}\sum_{i\in\sI} \frac{|\hat{\gU}_{i,N}\cap \gU_i|}{N},
% \end{equation}
% \begin{equation}
% \label{eq:ut_hr}
% \text{HitRate@}N = \frac{1}{|\sI|}\sum_{i\in\sI} \delta(|\hat{\gU}_{i,N}\cap \gU_i|>0),
% \end{equation}

% \begin{equation}
% \label{eq:ut_recall}
% \text{Recall@}N = \frac{1}{|\sI|}\sum_{i\in\sI} \frac{|\hat{\gU}_{i,N}\cap \gU_i|}{\text{Min}(|\gU_i|,N)},
% \end{equation}

\begin{table*}[!b]
\caption{Results of IR and UT obtained by the BCE loss with different negative sampling strategies versus the bbcNCE. The Metric is NDCG$@10$ for Amazon Books, Amazon Electronics, e\_comp and NDCG$@5$ for w\_comp. The best results are highlighted in bold, the second best is underlined. The \% is omitted.}
\label{tab:result_ce}
\centering
\begin{tabular}{ll|lll|lll|lll|lll}
\toprule
       &     & \multicolumn{3}{c|}{Amazon Books} & \multicolumn{3}{c|}{Amazon   Electronics} & \multicolumn{3}{c|}{QA e\_comp} & \multicolumn{3}{c}{QA w\_comp} \\
losses & NS: $p_n(u,i)$  & IR       & UT       & AVG      & IR          & UT         & AVG         & IR     & UT     & AVG     & IR     & UT     & AVG     \\ \midrule
BCE     & $\ptrain(u)$ & \underline{53.07} & 41.95 & 47.51 & \bftab24.43 & 10.50 & \underline{17.46} & \underline{36.99} & 4.98 & 20.98 & \underline{35.73} & 20.59 & 28.16 \\
BCE     & $\ptrain(i)$ & 42.85 & \underline{44.77} & 43.81 & 13.68 & 11.47 & 12.58 & 6.44  & \underline{7.35} & 6.90  & 24.29 & 22.24 & 23.27 \\
BCE     & $\ptrain(u)\ptrain(i)$ & 44.67 & 43.76 & 44.21 & 13.66 & \underline{11.81} & 12.73 & 6.08  & 6.41 & 6.25  & 24.78 & 21.57 & 23.17 \\
BCE     & $1/MK$ & 52.79 & 42.46 & \underline{47.63} & 24.34 & 9.81  & 17.08 & 36.51 & 6.70 & \underline{21.60} & 35.55 & \underline{23.42} & \underline{29.48} \\
bbcNCE & -  & \bftab57.20 & \bftab47.67 & \bftab52.44 & \underline{24.39} & \bftab12.77 & \bftab18.58 & \bftab37.65 & \bftab8.25 & \bftab22.95 & \bftab36.48 & \bftab24.30 & \bftab30.39 \\
\bottomrule
\end{tabular}
\end{table*}
\subsubsection{Experimental Comparisons}
%To do item recommendation and user targeting concurrently, our experiments compare various modeling strategies from different perspectives. 
First, we compare the results between modeling with the multinomial and Bernoulli distributions,
Then, the losses derived from modeling with the multinomial distribution are compared, and finally we study distinct model architectures. 
The detailed comparisons are listed as follows:
\begin{itemize}
\item \textbf{The bbcNCE versus the BCE.} Our proposed bbcNCE loss (Eq.~\ref{eq:j-infonce}) and the BCE loss with the specific negative sampling are the practical realizations of modeling the user-item interaction matrix with the multinomial and Bernoulli distributions, respectively. We experiment %with the bbcNCE and the BCE loss with various negative sampling methods, 
and evaluate their performance in both IR and UT of all the four datasets. %We fix the architecture to be the Youtube-DNN with mean pooling for a fair comparison.

\item \textbf{The bbcNCE versus other losses in the multinomial distribution scope.} When modeling the interaction matrix with the multinomial distribution, there are various implementations using different losses.
We compare the bbcNCE loss with other well-applied losses like SSM to show its effectiveness.% of the proposed bidirectional NCE loss with bias correction.
%For the SSM loss, we set the number of negative samples to be equal to the batch-size for a fair comparison with other losses. Among these losses, our proposed bbcNCE loss is superior to other well applied losses, as shown in Tab.~\ref{tab:optima-multinomial}. The results provide a strong support for our theoretical analysis in Sec. \ref{sec:loss}.

\item \textbf{The model-agnostic characteristic of the UniMatch framework.} 
%Our framework consists of the bbcNCE loss and a two-tower architecture. The bbcNCE loss is model agnostic, so 
We demonstrate that our framework is capable of integrating various models, including Youtube-DNN \cite{covington2016deep}, CNN used in Caser \cite{tang2018personalized}, RNN employed in GRU4Rec \cite{hidasi2015session} and Transformer utilized in SASRec \cite{kang2018self}, and show that they produce consistent results in the IR and UT tasks.

\item \textbf{The effectiveness of the incremental training.} We setup the training procedure as the incremental training month by month. By this way, the model can adapt to the latest distribution of user-item interactions, and produces much better results, particularly in the case that the item trends and users' interests shift quickly.

\item \textbf{Cost Saving.} %We summarize the various tactics to save the cost of applying IR and UT in QA, and show that we save up to 94\%+ of the total cost.
We summarize how the concurrent modeling of the IR and UT tasks by our framework saves the total cost up to 94\%+.
\end{itemize}

\subsubsection{Implementations}
%For a fair comparison, we use the popular Youtube DNN~\cite{covington2016deep} implemented by \cite{cen2020controllable} as the backbone architecture. Other complicated architectures can be easily incorporated into the UniMatch framework.

The code of our experiments is implemented with TensorFlow 1.12~\cite{abadi2016tensorflow} in Python 2.7, running on Nvidia GPU Tesla T4. We use the existing CNN and RNN modules in TensoFlow, and implement the Transformer based on this github repository\footnote{\url{https://github.com/Kyubyong/transformer}}.

\begin{table*}[t]
\caption{Results of IR and UT of bbcNCE loss versus other losses modeling $\vs_r$ or $\vs_c$ with the multinomial distribution on Amazon datasets. The $\%$ is omitted. The details of the losses can be found in Tab.~\ref{tab:optima-multinomial}. `SSM w. n' is the SSM loss with the users' and items' representations being l2-normalized. The best results are highlighted in bold, the second best result is underlined.}
\label{tab:result_amazon}
\centering
\begin{tabular}{l|llllll|llllll}
\toprule
                                    & \multicolumn{6}{c|}{Amazon Books}                                                           & \multicolumn{6}{c}{Amazon Electronics}                         \\
                                    & \multicolumn{2}{c}{IR}          & \multicolumn{2}{c}{UT}         & \multicolumn{2}{c|}{AVG} & \multicolumn{2}{c}{IR}         & \multicolumn{2}{c}{UT}  & \multicolumn{2}{c}{AVG}      \\
\midrule
\multicolumn{1}{l|}{loss}           & Recall         & NDCG           & Recall         & NDCG          & Recall         & NDCG    & Recall         & NDCG          & Recall        & NDCG   & Recall       & NDCG      \\ 
\midrule
SSM w. n.   & 77.07 & 56.88 & 58.01 & 35.85 & 67.54 & 46.36 & \underline{48.78} & \underline{25.82} & 13.14 & 6.06  & 30.96 & 15.94 \\
\midrule
InfoNCE     & 70.73 & 48.12 & 68.78 & 46.67 & 69.76 & 47.39 & 28.19 & 15.83 & \bftab24.54 & \bftab13.35 & 26.36 & 14.59 \\
SimCLR      & 71.53 & 49.35 & \underline{69.24} & 47.64 & 70.38 & 48.50 & 27.97 & 15.83 & \underline{22.96} & 12.68 & 25.46 & 14.26 \\
\midrule
row-bcNCE   & \bftab78.56 & \bftab58.71 & 66.71 & 44.44 & \underline{72.64} & \underline{51.58} & \bftab49.54 & \bftab28.88 & 20.13 & 11.00 & \bftab34.83 & \bftab19.94 \\
col-bcNCE   & 68.23 & 46.03 & \bftab71.12 & \bftab50.42 & 69.67 & 48.23 & 25.68 & 14.33 & 21.57 & 11.95 & 23.63 & 13.14 \\
bbcNCE    & \underline{77.43} & \underline{57.20} & 69.18 & \underline{47.67} & \bftab73.31 & \bftab52.44 & 41.89 & 24.39 & 22.55 & \underline{12.77} & \underline{32.22} & \underline{18.58} \\ 
\bottomrule
\end{tabular}
\end{table*}

\subsection{Experimental Results}
\label{sec:res}
%We show extensive comparison results, including 1). the bbcNCE loss versus the BCE loss; 2). the bbcNCE loss versus other losses in the multinomial distribution scope; 3). Various models trained with the bbcNCE loss.

\subsubsection{\textbf{The bbcNCE versus the BCE}}
\label{sec:vs-bce}
%The results of the losses are obtained with the their own optimal hyper-parameters. 
The backbone model is the Youtube-DNN with mean pooling for all the losses. As shown in Tab. \ref{tab:result_ce}, we have these observations:

$i).$ We compare different negative sampling methods with the BCE loss. Negative sampling with $p_n(u,i) \propto \ptrain(u)$ gives consistent good results in the IR, while $p_n(u,i) \propto \ptrain(i)$ performs well in the UT. The uniform sampling with $p_n(u,i)=1/MK$ gives equally good results for both the IR and UT. %This is in align with our theoretical analysis in Sec. \ref{sec:optimum-b}, as summarized in Tab. \ref{tab:optima-bernoulli}.

Particularly, the negative sampling with $\ptrain(u)$ outperforms $\ptrain(i)$ by $51.2\%$ on average for the IR task on the Amazon datasets. On our QA datasets, the difference is more significant. The NDCG@$10$ is almost 5 times higher on the QA e\_comp dataset.
In contrast, the results obtained from the negative sampling with $\ptrain(i)$ surpasses $\ptrain(u)$ for the UT task. On the QA e\_comp dataset, the metric result is about 48\% relatively higher. 

The above comparisons show that the IR and UT tasks require different sampling methods for the BCE loss to achieve competing results.
%This is in align with our theoretical analysis in Sec. \ref{sec:optimum-b} that these sampling methods guide the converged model output $\phi_\theta(u,i)$ to be proportional to $\log\ptrain(i|u)$, $\log\ptrain(u|i)$ and $\log\ptrain(u,i)$. 

$ii).$ The proposed bbcNCE loss obtains the best or second best results for both IR and UT across all the datasets. This verifies that in theory modeling the user-item interaction matrix $\mS_{ui}$ with Bernoulli and multinomial distributions makes no difference, but in practice the bbcNCE can reach better and robust results.% consistently. 

We hypothesize that it is due to the comparison mechanism rooted in the loss. In the BCE loss, the model parameter $\theta$ is optimized to push the sample's sigmoid towards 1 or 0. 
%This means the samples are relatively `isolated'\footnote{It is not totally `isolated' during the training, because it will compete with other samples on providing the gradients in the back propagation.} during the optimization since it only compares with 1 or 0. 
With the bbcNCE loss, the positive items/users are forced to compare with other items/users in the same batch as detailed in Tab. \ref{tab:infonce-loss}, and the model is optimized to allow the positive item/user to surpass all the others.

% \begin{table*}[t]
% \centering
% \caption{The simulated costs using two dummy datasets with different sizes. `m': million. The cost is calculated using the actual cloud service price on \textit{aliyun.com}, and the unit is Chinese Yuan.}
% \label{tab:cost}

% \begin{tabular}{l|l|ll|lll}
% \toprule
%               &               & \multicolumn{2}{c|}{prediction data} & \multicolumn{3}{c}{cost}                                                                \\
%               & training data & \# users       & \# items      & \multicolumn{1}{c}{train} & \multicolumn{1}{c}{prediction} & \multicolumn{1}{c}{total} \\
% \midrule
% middle-size & 5 m           & 1 m            & 100           & 9.77                       & 1.47                           & 11.24                     \\
% large-size  & 40 m          & 5 m            & 200           & 70.04                      & 1.95                           & 71.99                     \\
% \bottomrule
% \end{tabular}
% \end{table*}

$iii).$ The bbcNCE loss costs much less than the BCE loss during training. As illustrated in Tab. \ref{tab:hyper-param}, the losses derived from the multinomial modeling paradigm requires much less training epochs to converge. For the Amazon books dataset, the BCE loss reaches the best results with 8 epochs, while bbcNCE converges in 3 epochs. In addition, the BCE loss processes 2 times of data due to the $1:1$ sampling of negatives per epoch as illustrated in Tab. \ref{tab:infonce-loss} and \ref{tab:ce-loss}. Therefore, the computation cost of training is about 5 times. For other datasets, the computation cost is about $6\sim 10$ times, which is calculated from the epochs in Tab. \ref{tab:hyper-param}. 

We conjecture that the comparison mechanism stated above applies here: in each epoch, a sample in the BCE loss does not provide additional information except for its divergence from the true label 1 or 0. According to the information theory, it provides no more than 1 bit information. On the contrary, a sample in the bbcNCE loss is employed to differentiate one positive item/user from the rest of the items/users in the same batch as in Tab. \ref{tab:infonce-loss}, so it can offer at most $\log_{2}64=6$ bits of information if the batch-size is 64.
To conclude, the bbcNCE can utilize much more information per-epoch during the training, and thus speeds up the convergence and reduce the cost by a large amount.
% This verifies our theoretical proof of their equivalence in Sec. \ref{sec:prob-def}, and show that they can guide the models to converge to the same optima. We also show that the latter is more efficient during training, and also gives robust results. 

%Based on the above experiments, we choose to model the user-item interactions with the multinomial distribution in our \textit{UniMatch} framework, implemented with the bbcNCE loss.

\begin{table*}[t]
\caption{Results of IR and UT of the bbcNCE loss versus other losses modeling $\vs_r$ or $vs_u$ with the multinomial distribution on QuickAudience datasets. The best results are highlighted in bold, the second best is underlined. The \% is omitted.}
\label{tab:result_qa}
\centering
\begin{tabular}{l|llllll|llllll}
\toprule
                                    & \multicolumn{6}{c|}{QA e\_comp}                                                                & \multicolumn{6}{c}{QA w\_comp}                         \\
                                    & \multicolumn{2}{c}{IR}          & \multicolumn{2}{c}{UT}         & \multicolumn{2}{c|}{AVG} & \multicolumn{2}{c}{IR}         & \multicolumn{2}{c}{UT}  & \multicolumn{2}{c}{AVG}      \\
\midrule
\multicolumn{1}{l|}{loss}           & Recall         & NDCG           & Recall         & NDCG          & Recall         & NDCG    & Recall         & NDCG          & Recall        & NDCG   & Recall       & NDCG      \\ 
\midrule
SSM w. n.   & 58.22 & 35.57 & 15.37 & 6.85 & 36.79 & 21.21 & 49.17          & {\ul 36.54}    & 34.95          & 23.95          & {\ul 42.06}    & {\ul 30.25}    \\
\midrule
InfoNCE     & 15.82 & 7.33  & 14.62 & 7.29 & 15.22 & 7.31  & 38.14          & 28.63          & 27.59          & 18.09          & 32.87          & 23.36          \\
SimCLR      & 16.25 & 7.19  & 15.30 & 7.18 & 15.77 & 7.18  & 36.56          & 27.26          & 35.36          & 24.10          & 35.96          & 25.68          \\
\midrule
row-bcNCE   & \bftab62.37 & \bftab38.49 & 15.98 & 7.27 & \underline{39.17} & \underline{22.88} & \textbf{50.17} & \textbf{37.10} & 31.53          & 20.92          & 40.85          & 29.01          \\
col-bcNCE   & 25.62 & 12.61 & \bftab18.09 & \bftab8.35 & 21.86 & 10.48 & 34.32          & 24.24          & \textbf{38.42} & \textbf{24.87} & 36.37          & 24.56          \\
bbcNCE    & \underline{61.35} & \underline{37.65} & \underline{17.63} & \underline{8.25} & \bftab39.49 & \bftab22.95 & {\ul 49.54}    & 36.48          & {\ul 35.47}    & {\ul 24.30}    & \textbf{42.50} & \textbf{30.39} \\
\bottomrule
\end{tabular}
\end{table*}

\begin{table*}[t]
\caption{Statistics of the popularity/activeness of items/users retrieved by different losses. We measure the median and average of items' popularity and users' activeness. The popularity and activeness are defined as the number of interactions occurred in the past one year. Here `med' and `avg' stands for median and average respectively.}
\label{tab:retrieved_obj}
\centering
\begin{tabular}{l|llll|llll|llll|llll}
\toprule
          & \multicolumn{4}{c|}{Amazon Books}                                                                        & \multicolumn{4}{c|}{Amazon   Electronics}                                                                & \multicolumn{4}{c|}{QA e\_comp}                                                                             & \multicolumn{4}{c}{QA w\_comp}                                                                             \\
          & \multicolumn{2}{c}{IR}                            & \multicolumn{2}{c|}{UT}                            & \multicolumn{2}{c}{IR}                            & \multicolumn{2}{c|}{UT}                            & \multicolumn{2}{c}{IR}                            & \multicolumn{2}{c|}{UT}                            & \multicolumn{2}{c}{IR}                            & \multicolumn{2}{c}{UT}                            \\
losses    & med       & avg      & med      & avg       & med       & avg      & med       & avg      & med       & avg      & med       & avg      & med      & avg       & med       & avg      \\
\midrule
SSM w. n.      & 25           & 72       & 6           & 13.6      & 232          & 491      & 4            & 4.8      & 94           & 187      & 4            & 6.4      & 11969       & 15176     & 5            & 7.1      \\
\midrule
InfoNCE   & 16           & 46       & 6           & 13.6      & 34           & 139      & 4            & 5.1      & 52           & 104      & 4            & 6.3      & 2332        & 5815      & 4            & 6.3      \\
SimCLR    & 16           & 47       & 6           & 13.3      & 33           & 125      & 4            & 5.1      & 55           & 113      & 4            & 6.5      & 2294        & 5961      & 5            & 7.1      \\
\midrule
row-bcNCE & 27           & 78       & 6           & 12.9      & 236          & 496      & 4            & 4.9      & 138          & 245      & 4            & 6.6      & 11969       & 15189     & 5            & 6.7      \\
col-bcNCE & 16           & 47       & 6           & 14.7      & 39           & 150      & 4            & 5.2      & 96           & 173      & 4            & 7.2      & 3063        & 6456      & 5            & 7.8      \\
bbcNCE    & 23           & 69       & 6           & 14.1      & 160          & 400      & 4            & 5.1      & 138          & 246      & 5            & 7.4      & 12837       & 15320     & 5            & 7.1     \\
\bottomrule
\end{tabular}
\end{table*}

\subsubsection{\textbf{The bbcNCE versus other losses in the multinomial distribution scope}}
\label{sec:vs-multinomial}
%As stated in Sec.\ref{sec:unimatch}, our \textit{UniMatch} framework employs bbcNCE loss to model $\vs_r$ and $\vs_c$ concurrently. 
In comparison with other losses that modeling either $\vs_r$ or $\vs_c$ or both, the bbcNCE guides the model to learn the joint probability $p(u,i)$ from the empirical distribution $\ptrain(u,i)$. 
The experimental results in Tab.~\ref{tab:result_amazon} and \ref{tab:result_qa} are in alignment with our theoretical proof. 
We analyze the experiments from the following three aspects.

$i).$ We can use only one model trained with the bbcNCE loss to serve for both IR and UT in our applications. For IR, the bbcNCE is on par with the losses that model $\vs_r$ and lead to the convergence at $\ptrain(i|u)$, \ie, SSM and row-bcNCE. For UT, its results match with the col-bcNCE loss that models $\vs_c$ and converges at $\ptrain(u|i)$ as in Tab. \ref{tab:optima-multinomial}. For both IR and UT, the bbcNCE loss produces the best or second best results robustly, which makes it the best choice for our QA applications.

$ii).$ The bias correction plays the key role on lifting the results in IR. As shown in Tab. \ref{tab:result_amazon} and \ref{tab:result_qa}, the bbcNCE and row-bcNCE losses always produce much better results than the InfoNCE and SimCLR losses that have no bias correction. 
The SSM loss is implemented with negative sampling and bias correction so that it converges to $\ptrain(i|u)$ in theory. However, its performance is usually inferior than the bbcNCE and row-bcNCE. 
The SSM loss draws negative samples from the whole item vocabulary, while in contrast the bbcNCE samples negatives from in-batch training data. 
Therefore, in our monthly incremental training setting, the bbcNCE only encounters negative items within the current month, and the SSM compares the positive items to the negative items sampled from the whole vocabulary. This brings the advantage of the bbcNCE on fitting on the latest data distribution, thus makes the results better.

In UT, the performance of bbcNCE and col-bcNCE does not always surpass other losses by a large margin. 
This implies that the bias correction from the users' empirical distribution is not that effective anymore when the data is sparse.
We speculate that the marginal distribution calculated from the sparse dataset is not reliable. Because most users only have very few interactions, the $\ptrain(u)$ computed will not be statistically significant.
%For example, the Amazon Electronics dataset as in Tab. \ref{tab:exp_stats}.
On the contrary, if the user behavior data is relatively rich, the col-bcNCE and bbcNCE that correct the users' distribution bias produce much better results compared to the InfoNCE and SimCLR loss, as shown on the Amazon Books, QA e\_comp and w\_comp datasets.
%As shown in Tab. \ref{tab:result_amazon} and \ref{tab:result_qa}, in the Amazon Books, QA e\_comp and w\_comp datasets, 

$iii).$ The experiment results in Tab.~\ref{tab:result_amazon} and \ref{tab:result_qa} show that the SimCLR and InfoNCE losses yield very close results. This is in align with our claim that they both converges at $\ptrain(u,i)/(\ptrain(u)\ptrain(i))$ as in Tab. \ref{tab:optima-multinomial}. As shown in \cite{oord2018representation}, the InfoNCE loss converges when the mutual information between the two matching variables is maximized. When it is applied in IR, it will tend to recommend users with unpopular items, because those items usually have high point-wise mutual information with the users.

We use the number of historical interactions to measure the popularity/activeness of items/users as in Tab. \ref{tab:retrieved_obj}. For example, if an item is purchased 100 times in the past 1 year, then its popularity is 100. For all the top-$n$ items/users retrieved by the model, we calculate the median and average popularity/activeness. 
Taking the Amazon Books dataset as an example, the median item popularity from the InfoNCE and SimCLR losses is 16, which means that 50\% of the recommended items have less or equal than 16 interactions in the past 1 year. In contrast, the median popularity of the bbcNCE, row-bcNCE and SSM losses ranges from 23 to 27. In all the datasets in Tab. \ref{tab:retrieved_obj}, the InfoNCE and SimCLR losses always tend to recommend unpopular items. In UT, the two losses prefer inactive users in general, but it is not that evident because most users do not have many interactions.

%The results are coherent with the theory, but it also suggests that the InfoNCE or SimCLR losses are improper for IR and UT, particularly IR, as shown in Tab. \ref{tab:result_amazon} and \ref{tab:result_qa}.

\begin{table*}[t]
\caption{Results of IR and UT of bbcNCE loss versus other losses evaluated using various models on our QA w\_comp datasets. The metric is NDCG$@5$. Each column represents one kind of model, and `CEX' means the context extractor of the model, and `AGG' is the sequence aggregator. `mean', `last', `attn' are mean, last and attention pooling respectively. The best results are highlighted in bold, the second best is underlined. The \% is omitted.}
\label{tab:model_agnostic}
\centering
\begin{tabular}{ll|lll|lll|lll|lll|lll}
\toprule
 &CEX & \multicolumn{3}{c|}{Youtube-DNN} & \multicolumn{3}{c|}{CNN-$l1$} & \multicolumn{3}{c|}{GRU} & \multicolumn{3}{c|}{LSTM} & \multicolumn{3}{c}{Transformer-$l1$} \\
 &AGG & mean    & last    & attn    & mean    & last    & attn   & mean   & last   & attn  & mean   & last   & attn   & mean       & last      & attn      \\

\midrule
\multirow{6}{*}{IR} & SSM w. n. & {\ul 36.54}    & 27.51          & {\ul 36.62}    & 35.67          & 30.61          & 36.05          & {\ul 36.85}    & 35.21          & \textbf{36.40} & 35.93          & {\ul 35.92}    & 35.59          & 35.27          & \textbf{34.89} & \textbf{37.06} \\
% \midrule
\cmidrule{2-17}
 & InfoNCE   & 28.63          & 19.84          & 27.58          & 27.82          & 23.38          & 27.12          & 28.13          & 27.58          & 27.12          & 27.12          & 27.98          & 28.39          & 27.91          & 25.67          & 26.22          \\
 & SimCLR    & 27.26          & 20.45          & 27.17          & 27.48          & 23.96          & 26.59          & 27.46          & 27.32          & 26.95          & 26.47          & 26.89          & 27.49          & 26.02          & 24.92          & 25.87          \\
% \midrule
\cmidrule{2-17}
 & row-bcNCE & \textbf{37.10} & {\ul 28.35}    & {\ul 36.61}    & \textbf{36.86} & {\ul 31.08}    & {\ul 37.01}    & 36.27          & {\ul 36.22}    & {\ul 35.87}    & \textbf{36.69} & \textbf{36.47} & \textbf{36.53} & \textbf{36.81} & 33.95          & {\ul 36.79}    \\
 & col-bcNCE & 24.24          & 9.10           & 26.25          & 24.76          & 11.66          & 23.52          & 19.93          & 19.02          & 18.80          & 16.90          & 12.72          & 16.58          & 15.89          & 17.73          & 24.99          \\
 & bbcNCE    & {\ul 36.48}    & \textbf{28.52} & \textbf{36.77} & {\ul 36.26}    & \textbf{31.39} & \textbf{37.12} & \textbf{37.05} & \textbf{36.31} & 35.70          & {\ul 36.11}    & 35.85          & {\ul 35.98}    & {\ul 36.02}    & {\ul 34.78}    & 36.43          \\
\specialrule{.1em}{.3em}{.3em}
% \midrule
\multirow{6}{*}{UT} & SSM w. n. & 23.95          & 13.61          & 23.88          & 25.17          & 14.97          & 23.65          & 22.01          & 24.49          & 25.61          & 24.49          & {\ul 26.02}    & 24.02          & 20.53          & 20.15          & 23.25          \\
% \midrule
\cmidrule{2-17}
 & InfoNCE   & 18.09          & {\ul 13.91}    & 19.92          & 24.21          & 13.35          & 22.85          & 23.91          & 23.78          & 24.08          & 26.24          & 24.07          & 24.31          & 19.21          & 19.66          & 20.24          \\
 & SimCLR    & 24.10          & 13.73          & 23.57          & 23.89          & 14.86          & 25.12          & 25.72          & 26.23          & 25.43          & {\ul 26.41}    & 24.54          & 26.01          & \textbf{25.68} & 20.69          & 23.40          \\
% \midrule
\cmidrule{2-17}
 & row-bcNCE & 20.92          & 13.63          & 20.61          & 23.04          & 16.86          & 22.42          & 24.66          & 24.39          & 24.96          & 23.89          & 25.09          & 25.09          & 22.33          & 20.23          & 21.09          \\
 & col-bcNCE & \textbf{24.87} & \textbf{14.64} & \textbf{26.54} & \textbf{26.53} & \textbf{18.65} & {\ul 26.72}    & {\ul 26.82}    & {\ul 28.50}    & {\ul 25.81}    & 25.79          & 24.44          & {\ul 28.02}    & 24.03          & \textbf{24.02} & \textbf{25.93} \\
 & bbcNCE    & {\ul 24.30}    & 13.33          & {\ul 24.29}    & {\ul 25.67}    & {\ul 18.35}    & \textbf{26.79} & \textbf{27.64} & \textbf{29.47} & \textbf{26.63} & \textbf{27.85} & \textbf{27.93} & \textbf{28.88} & {\ul 25.44}    & {\ul 22.64}    & {\ul 24.06}   \\
\bottomrule
\end{tabular}
\end{table*}

\subsubsection{\textbf{The model-agnostic characteristic of the UniMatch
framework}}
\label{sec:agnostic}
%Our \textit{UniMatch} framework can incorporate various models as the user or item encoders. 
We implement 5 types of context extractors, \ie, Youtube-DNN\footnote{Youtube-DNN represents no context extractor here, \ie, the lookup item embeddings are directly passed to the aggregation layer.}, 1-layer CNN, GRU, LSTM, and 1-layer Transformer, and 4 types of aggregators, \ie, mean pooling, last pooling, max pooling and attention pooling. 
So we have 20 models in total. 
We report the results of QA w\_comp dataset in Tab. \ref{tab:model_agnostic}. 
The results of Max pooling are always worse than other aggregators, so we omit them. 

%The results of Youtube-DNN or CNN-$l1$ with last pooling yields inferior results, because Last pooling cannot utilize all the information of the input sequence due to the limited receptive field. Transformer-$l1$ with last pooling does not produce good results either. This indicates that a shallow transformer cannot extract and summarize the contextual information into the vector of a particular position, \eg, the last vector of the sequence. On the contrary, GRU and LSTM can encapsulate the related information in the last vector, so that results of last pooling can match or outperform the results from mean or attention pooling.

In general, the results from different models under the same loss does not differ notably. This suggests that the complicated context extraction methods that thrive in processing the text sequence are not superior when dealing with users' sequential behavior data, \eg, the Amazon and QA datasets. 
It further implies that the contextual information is not very important on understanding users' single interaction with an item. 
In contrast, a word's exact meaning is defined by its context in natural language processing (NLP). 
This finding motivates us to adopt the Youtube-DNN with mean pooling as our default model for QA to save the computation cost in practice.

Across all the models, the proposed bbcNCE loss gives best or close to the best results in both tasks. For IR, the bbcNCE and row-bcNCE are the top two in almost all the models, while the bbcNCE and col-bcNCE are the best for UT. 
The results further verify our statements on the losses' optima as in Tab. \ref{tab:optima-multinomial}, regardless of the choice of models.

\begin{figure*}[t]
  \begin{center}
  \includegraphics[width=0.8\textwidth]{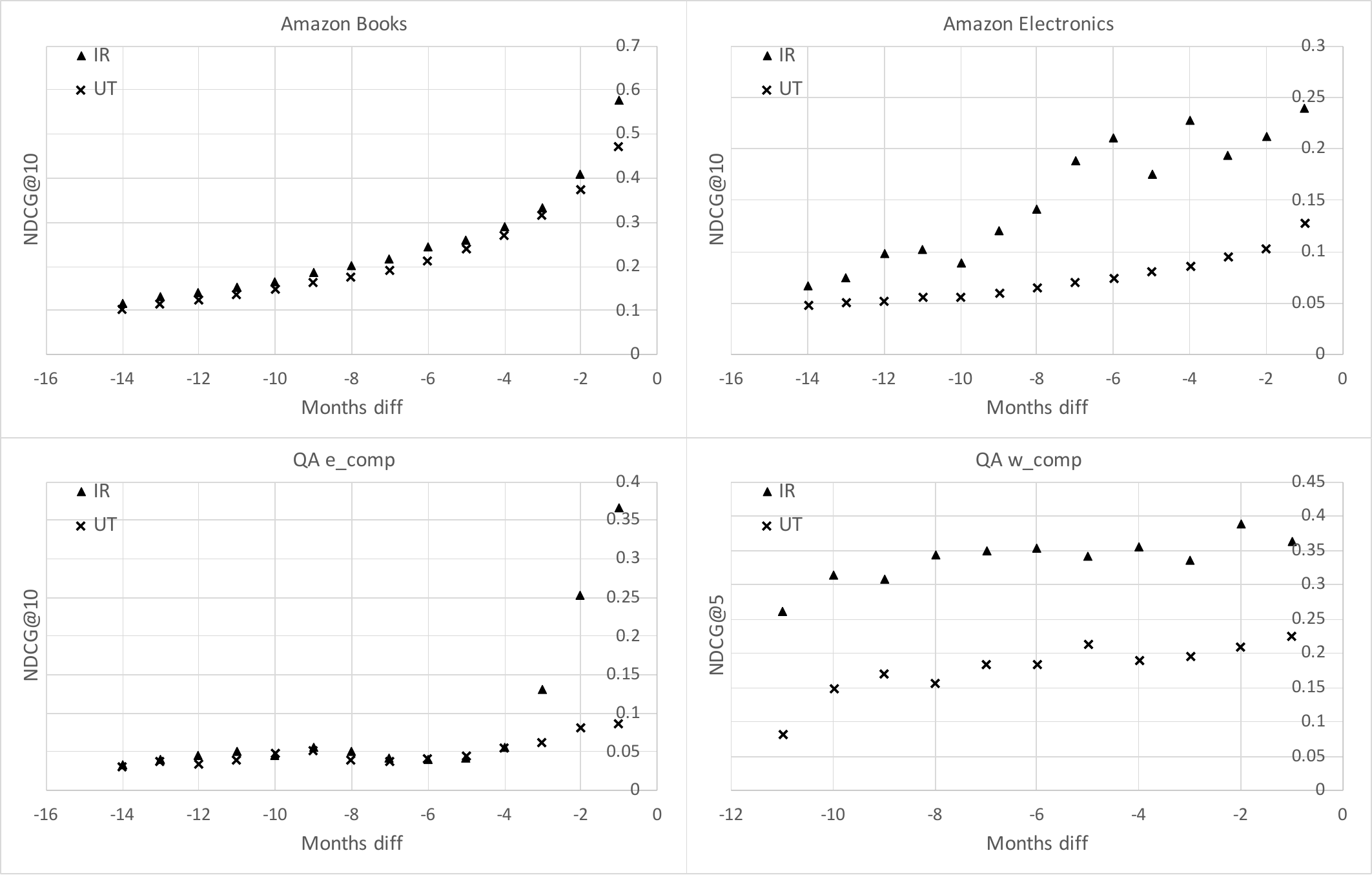}
  \end{center}
  \caption{The intermediate results from the incremental training setup for the Amazon and QA datasets. The $y$-axis is the NDCG metric, and $x$-axis represents the number of months ahead of the date of the test data. The metric increases steadily as newer data is feed into training.}
  \label{fig:incremental}
\end{figure*}
\subsubsection{\textbf{The effectiveness of the incremental training}}
\label{sec:incremental}
In our \textit{UniMatch} framework, we train the model incrementally month by month. We adopt this training setup for our specific applications in QA, because IR and UT campaigns are usually organized monthly in private domain marketing. 
In addition, the incremental training can save cost, and improve the prediction results as depicted in Fig. \ref{fig:incremental}.

We observe that the gain of the incremental training is crucial on the Amazon Books and QA e\_comp datasets. The NDCG metric of the model trained till 1 months before the test date is much higher than trained till 2 or 3 months before. We speculate that their items are very sensitive to time. For example, users prefer to buy new and popular books, which may vary quickly. So we have to keep training the model with the latest data to adapt for the trend. On the other hand, the results of the Amazon Electronics and QA w\_comp datasets are relatively stable during the incremental training. This implies their items are more stable over time, and their users' interests are relatively static.

\subsubsection{\textbf{Cost Saving}}

% We estimate the cost using dummy data, as shown in Tab.~\ref{tab:cost}. 
%The main cost of the application in QA is the GPU training cost. We have employed 4 tactics to reduce the cost as follows:
We demonstrate that the flexibility of our framework and the proposed bbcNCE loss enable a significant cost saving in practice.

$i).$ We choose the bbcNCE instead of the BCE loss to reduce the data consumption during the training as in Sec. \ref{sec:vs-bce}. Thus we reduce the cost to $1/10 \sim 1/5$. Our theoretical analysis and experimental results show that the performance is on par or better.

$ii).$ We propose the bbcNCE so that we can train only one model to do both IR and UT without performance decline as depicted in Sec. \ref{sec:vs-multinomial}. We can reduce both the training and prediction cost to $1/2$. At the same time, the underlying management cost of multiple models is also saved.

$iii).$ We analyse various popular models and choose the simplest Youtube-DNN with mean pooling as our default model. Experiments in Sec. \ref{sec:agnostic} show that it can generate SOTA comparable results on all the datasets. 
Therefore, we do not have to choose too complex models and relieve ourselves from the large computation burden.
%we get rid of the large computation burden of the complex models.

$iv).$ We adopt the incremental training setup as illustrated in Sec. \ref{sec:incremental}. With the conventional training strategy, we use past 1 year data to train from scratch monthly. Using the incremental training, we can just utilize the past 1 month data and train from the latest model checkpoint. By this way, we can reduce the training cost to $1/12$.

% In reality, merchants need to do both item recommendation and user targeting. As our \textit{UniMatch} framework can do the job with just one model, thus cut the cost by half.
%Assume we train the model every month to adapt for the latest data. 

To conclude, in the applications of QA, our \textit{UniMatch} framework can reduce the training cost up to $1/240 \sim 1/120$ while keeping the SOTA comparable performance. The prediction cost is reduced to 1/2 and eliminate the management cost of multiple models. The training cost is usually about 90\% of total cost, so we can save up to 94\%+.

% modeling the point-wise mutual information (PMI)

% \subsubsection{Effects of Temperature and Batch Size}
% [place holder: plot temperature \& batchsize versus performance] 
\section{Related Work}
\label{sec:related}

\subsection{Item Recommendation}
Item recommendation (IR) have been studied in both academia and industry for decades. The collaborative filtering (CF) and its variants are widely adopted in the early years \cite{su2009survey}. 
Later its descendant, the matrix factorization (MF) \cite{funk2006}, is proposed to solve the problem more elegantly with higher accuracy. 
Then, the Probabilistic Matrix Factorization (PMF) \cite{mnih2008probabilistic} builds a solid theoretic foundation for the MF models based on the probability theory, \ie, PMF models $s$ with Gaussian distributions. Later, the Bernoulli distribution is shown to be superior in modeling $s$ \cite{he2017neural}.

%Recently, the neural networks are adopted in the IR~\cite{he2017neural,rendle2020neural}. 
In recent years, the neural networks have become a significant component for the recommendation algorithms~\cite{he2017neural,rendle2020neural}, and contributed greatly for the recommender systems in industry. 
There are two common stages in an large-scale industrial recommendation application, \ie, the candidate generation stage and the ranking stage. % This practice is aim to balance the efficiency and accuracy in online serving.
The former stage is usually formulated as a multi-class classification problem to quickly select a small set of item candidates from a vast number of items \cite{covington2016deep,li2019multi,cen2020controllable}. 
%This stage is crucial for balancing the efficiency and accuracy in online serving \cite{covington2016deep,li2019multi,cen2020controllable}.
%In the ranking stage, the problem is formulated as the binary classification, and various models has been developed: wide \& deep model \cite{cheng2016wide} captures explicit and implicit feature crossing, Deep Interest Network \cite{zhou2018deep} and Deep User Perception Network \cite{ni2018} utilize attention mechanism \cite{bahdanau2014neural} to capture users' short-term interests.
In the ranking stage, the problem is formulated as a binary classification to rank all the selected candidates \cite{cheng2016wide,zhou2018deep,ni2018}.
% There are various influential models which have been well used in real-world applications.
% For instance, the Wide \& Deep model captures explicit and implicit feature crossing \cite{cheng2016wide}. 
% The Deep Interest Network \cite{zhou2018deep} and Deep User Perception Network \cite{ni2018} utilize the attention mechanism \cite{bahdanau2014neural} to capture users' short-term interests.
Although not directly declared in many of these research, the underlying probability theory for the candidate generation stage is to model $\vs_r$ with the multinomial distribution \cite{liang2018variational}, and is to model $s$ with the Bernoulli distribution for the ranking stage \cite{he2017neural}.

In the candidate generation stage, the huge number of items causes problems on calculating the partition function of the loss during the optimisation (as in Eq. \ref{eq:u2i-loss}). The sampled softmax (SSM) loss \cite{jean2015using} is widely employed to solve the problem \cite{covington2016deep,li2019multi}. Recently, the InfoNCE loss is exploited in item recommendation to suppress popular items during candidate generation \cite{zhou2021contrastive}.
% It can be easily implemented with in-batch negative sampling, and the features of $y$ can be fully utilized as well \cite{zhou2021contrastive}.

\subsection{User Targeting}
User targeting (UT) mines the potential users for given items. 
The \emph{item} could be anything that users can interact with, \eg, an insurance product \cite{kim2004intelligent}, a company/business~\cite{pennacchiotti2011democrats,pang2013simple,lo2016ranking}, a specific message (e.g., tweets) on social medias \cite{tang2015locating,gui2019mention} and even another user \cite{guy2018people}, etc. 
UT is usually formulated as a binary classification problem, and solved with models like SVM, LR and neural networks, etc \cite{hosmer2013applied,chang2011libsvm,chen2016}.
%These problems are usually treated as binary classification and solved with tools like SVM, LR and etc. for each individual item independently \cite{hosmer2013applied,chang2011libsvm,chen2016}.

In an e-commerce company, the \emph{item} could be a product, a brand, a product category, and a merchant, etc. The number of the \emph{items} ranges from thousands to hundreds of millions. It is impractical to model each item respectively, so we commonly model the items all together via binary classification like \cite{rendle2020neural}.

In the above applications, researchers implicitly model $s$ with the Bernoulli distribution, and the negative samples are generated with probability $p_n(u,i) \propto \ptrain(i)$.

\section{Conclusions}
In this work, we propose the \textit{UniMatch} framework that can be applied in both IR and UT to help merchants conduct the private domain marketing.
Our framework is model agnostic and consists of a two-tower architecture as well as the incremental training setup and the proposed bbcNCE loss. Through comprehensive comparisons with different losses, models and training procedures, we show that our framework can generate SOTA comparable results theoretically and experimentally. Our framework reduces more than 94\% of the total cost, making it affordable for merchants to do private domain marketing with the SOTA performance.

\bibliographystyle{splncs04}
\bibliography{icde_main}

\end{document}